\documentclass[12pt]{article}
\usepackage{graphicx}
\usepackage{amsmath,amssymb,amsfonts}
\usepackage{amsbsy}
\usepackage{array}
\usepackage{setspace}
\usepackage{float} 
\usepackage{subfig}
\usepackage{longtable,booktabs}
\usepackage{multirow} 
\usepackage{colortbl}
\usepackage{url}
\usepackage{tikz}
\usepackage[colorlinks, citecolor=blue,urlcolor=black]{hyperref}  
\usepackage{ulem}
\usepackage{cases}
\usepackage[numbers, square, sort&compress]{natbib}
\usepackage{tikz}
\usepackage{algorithm}
\usepackage{algpseudocode}
\usepackage{amsmath}

\usetikzlibrary{decorations.pathreplacing}
\usetikzlibrary{shapes.geometric, arrows,positioning}
\pgfdeclarelayer{background layer}
\pgfdeclarelayer{foreground layer}
\pgfsetlayers{background layer,main,foreground layer}
\usetikzlibrary{arrows.meta}
\def\@currentColor{white}
\tikzset{set color/.code = { \gdef\@currentColor{#1} },
        arrow/.style = {arrows={-Triangle[angle=35:2.5pt 6] }, line width=1pt}}

\renewcommand{\emph}[1]{\textit{#1}}
\newcounter{parnum}

\date{}
\begin{document}

        \title{A Novel Framework for Modeling Quarantinable Disease Transmission
}
\author{Wenchen Liu, Chang Liu, Dehui Wang and Yiyuan She}

\maketitle
\begin{abstract}
The COVID-19 pandemic has significantly challenged traditional epidemiological models due to factors  such as delayed diagnosis, asymptomatic transmission, isolation-induced contact changes, and underreported mortality. In response to these complexities, this paper introduces a novel CURNDS model   prioritizing compartments and transmissions based on contact levels, rather than merely on symptomatic severity or hospitalization status.
 The framework surpasses conventional uniform mixing and static rate assumptions by incorporating adaptive power laws, dynamic transmission rates, and spline-based smoothing techniques.

 The CURNDS model provides accurate estimates of undetected infections and undocumented deaths from COVID-19 data, uncovering the pandemic's true impact. Our analysis challenges the assumption of  homogeneous mixing between infected and non-infected individuals in traditional epidemiological models.
By capturing the nuanced transmission dynamics of infection and confirmation, our model offers new  insights into the spread of different COVID-19 strains. Overall, CURNDS provides a   robust framework for understanding   the complex transmission  patterns of highly contagious, quarantinable diseases.
\end{abstract}

\section{Introduction}
As of August 2024, the COVID-19 pandemic has caused unprecedented global disruption, with the World Health Organization (WHO) reporting over 790 million confirmed cases and more than 7.1 million deaths. The virus's rapid spread, along with frequent mutations such as the Alpha, Beta, and Omicron variants, has posed significant challenges to public health systems worldwide. Researchers have developed various types of models, each designed to capture different aspects of disease transmission and progression.

(i) \textbf{Compartmental models.} Compartmental models serve as a fundamental tool in understanding the spread of infectious diseases.  The Susceptible-Infected-Recovered (SIR) model \cite{Kermack1927} and its extension, the Susceptible-Exposed-Infected-Recovered (SEIR) model \cite{hethcote1976qualitative}, are extensively  used. The latter, particularly relevant in COVID-19 studies, includes an ``exposed'' category to account for  the incubation period. However, the assumption that the exposed individuals are entirely non-infectious contradicts observations that COVID-19 can be transmitted before symptoms manifest \cite{he2020temporal, ferretti2020quantifying}.

Further developments in modeling, such as   \cite{di2020dynamical,biswas2020modelling, yi2022characterizing, zhang2020prediction}, try to distinguish between \emph{symptomatic} and \emph{asymptomatic} cases  with different transmission rates. Despite these advancements, the  WHO  points out that transmission primarily occurs through \textbf{close contact} \cite{sayampanathan2021infectivity}.  Therefore, a model that prioritizes the ``level of contact'', over symptomatic presence, severity of symptoms, or hospitalization status \citep{keller2022tracking,yang2020modified,mwalili2020seir,hassan2023mathematical}, would provide a more precise and effective framework for understanding transmission and progression mechanisms.

(ii) \textbf{Univariate Time Series Forecasting.} This category of forecasting methods  utilizes only the historical data from a \emph{single} target time series. Traditional time series models like ARIMA are commonly employed \cite{alzahrani2020forecasting, abuhasel2022analyzing}, alongside a range of other statistical methods including exponential smoothing, support vector regression, and LASSO \cite{Rustam2020,bezerra2020prediction}.

Compared to the more comprehensive epidemiological compartmental models, these univariate methods---which rely solely on a single sequence of historical data---inherently lack the capability to capture the dynamic \emph{interactions} among various epidemiological groups, such as susceptible and infected populations.  However, despite this limitation, they often achieve high accuracy in forecasting short-term trends largely due to their simplicity and robustness \cite{bezerra2020prediction, ribeiro2020short}.


(iii) \textbf{Comprehensive Multivariate Forecasters.} This class of methods use a broad array of data inputs beyond just the target sequence \cite{mishra2024multivariate,somyanonthanakul2022forecasting, Silva2020}. Employing sophisticated machine learning techniques, such as deep neural networks and boosted trees, they excel in capturing highly non-linear relationships and demonstrate strong  predictive performance \cite{xu2022forecasting,alali2022proficient}.

Unlike compartmental models, which rely on  \emph{bilinear} relationships from {uniform} population mixing, or traditional \emph{linear} time series models, these approaches avoid rigid parametric assumptions.  On the other hand, their multivariate and nonparametric characteristics also incur high computational costs and can lead to model instability \cite{justus2018predicting}. Furthermore, these ``black-box" methods lack transparency \cite{wu2021interpretable}, hindering the interpretation of disease transmission mechanisms. A more balanced approach would integrate the  interpretability of compartmental models with the flexibility  of nonparametric techniques.
\\

This paper introduces the \textbf{CURNDS} model, a novel approach that incorporates new compartments and dynamic transmission mechanisms, moving beyond the limitations of traditional uniform mixing and static rate assumptions to more accurately estimate undetected infections and unreported deaths.
 The main contributions are as follows:

  \begin{enumerate}
    \item The CURNDS model  introduces new compartments, particularly for individuals unaware of their infection status who may not receive timely diagnoses. These individuals may eventually be confirmed through testing, recover without medical intervention, or pass away without official recognition \cite{zheng2021health}. The transmission processes, driven by {contact levels}, provide a comprehensive outline of the infection's trajectory, from initial exposure through disease progression to eventual resolution.

    \item The CURNDS model employs adaptive power transforms and time-varying transmission rates to enhance the depiction of initial infection and disease progression. This modification is essential, as the uniform population mixing assumption \cite{anderson1991}, fundamental to bilinear SIR-like compartmental models, does not adequately capture the complexities of the pandemic \cite{kwuimy2020nonlinear,Berx2022Epid}.   In fact, our analysis reveals that quarantine measures and other public health interventions drastically reduce contact between infected and non-infected individuals, leading to \textbf{sublinear} transmission dynamics.

    \item A critical component of our model is the incorporation of smoothing techniques to characterize time-varying nonparametric rate sequences. This method, based on natural cubic splines, not only reduces the number of free parameters but also boosts prediction accuracy through the use of an efficient and robust algorithm.

\item  When applied to real-world data, our method uncovers new insights into various aspects of the pandemic's dynamics, such as imbalanced power parameters, distinct strain behaviors, the effects of vaccination on reproduction numbers, and the under-reporting of active cases and uncertified deaths. These findings collectively highlight a significant underestimation of the pandemic's true impact.
\end{enumerate}

The remainder of this paper is organized as follows: We first detail  the formulation of the CURNDS model, including the categorization of the population and the   framework for transmission dynamics. Next, we discuss the statistical modeling and smoothing techniques used for parameter estimation, followed by a description of the computational methods employed to optimize the model's performance. We then present a case study using COVID-19 data from Quebec, Canada, showcasing the model’s application and predictive capabilities. The paper also offers a  comparative analysis between the wild-type strain and the Omicron variant. Finally, we conclude with   a summary   of the paper.

   \section{The CURNDS Model}\label{Sec_CURNDS Model}
   Dynamic compartment models are widely applied to model the evolution of epidemics. However, these models face challenges when dealing with diseases like COVID-19, which exhibit an infectious incubation period. Traditional models, such as the SEIR model, often assume that \textit{all} infected individuals have already been clinically confirmed, which is not the case with COVID-19. Infected individuals may \textit{not} be diagnosed timely, due to mild or even no symptoms during incubation period or lack of testing resources. In fact, many are not aware of their infectious status, but  may unfortunately be highly contagious. Similarly, the infected individuals may have healed or died \textit{before} receiving confirmation through diagnostic testing. Therefore, it is necessary to define some new groups (such as $\boldsymbol{U}$, $\boldsymbol{S}$ in Figure \ref{Fig_sixgroups}) and new transmissions (cf. Figure \ref{Fig_U}) to capture the unique characteristics of quarantinable diseases like COVID-19.
   \subsection{\{$\boldsymbol{N}$, $\boldsymbol{S}$, $\boldsymbol{R}$\} vs. \{$\boldsymbol{U}$, $\boldsymbol{C}$, $\boldsymbol{D}$\}}\label{Sec_Model Composition}
    \begin{figure}[htbp]
        \centering
        \tikzstyle{groupbox} = [rectangle, rounded corners, minimum width = 1.5cm, minimum height=1cm,text width=2cm,text centered, draw = black]
        \begin{tikzpicture}[node distance=2.0cm,auto,>=latex',thin]
                \node[groupbox, draw = black, fill = purple!30](S){\textbf{S}\scriptsize elf-healed individuals {\fontsize{4}{5}\selectfont\mbox{(\textcolor{blue}{\textbf{\textit{unaware\hspace{-0.4em} of\hspace{-0.4em} infection}}})}}};
                \node[groupbox, below of = S,yshift=-1.4cm,draw = black, fill =purple!30](R){\textbf{R}\scriptsize ecovered individuals {\fontsize{4}{4.5}\selectfont\mbox{(\textcolor{blue}{\textbf{\textit{after\hspace{-0.4em} testing\hspace{-0.4em} positive}}})}}};
                \node[groupbox, right of = S, xshift = 4cm,draw = black, fill = yellow!40](U){\textbf{U}\scriptsize naware infections};
                \node[groupbox, above of = U,yshift=1.4cm,draw = black, fill = green!40](N){\textbf{N}\scriptsize on-infected individuals};
                \node[groupbox, below of = U, yshift = -1.4cm,draw = black,fill = orange!40](C){\textbf{C}\scriptsize onfirmed cases};
                \node[groupbox, right of = C, xshift = 4cm,draw = black, fill =gray!40](D){\textbf{D}\scriptsize eceased individuals};
                \draw [arrow,ultra thick] (N) -- node[]{} (U) ;
                \draw [arrow,thin] (U) -- node[]{}(S);
                \draw [->, thin] (U) -| node[xshift=-3cm]{} (D) node[pos=0.95,left] {\scriptsize $\boldsymbol{D}_{U}$};
                \draw [arrow,ultra thick] (U) -- node[xshift=-0.6cm]{}(C);
                \draw [arrow,thin] (C) -- node[]{}(R);
                \draw [arrow,thin] (C) --node[pos=0.9,above]{\scriptsize $\boldsymbol{D}_{C}$} (D) ;
                \draw [dashed,color=black] (-1.3,4.1)--(7.3,4.1);
                \draw [dashed,color=black] (7.3,4.1)--(7.3,2.75);
                \draw [dashed,color=black] (1.3,2.75)--(7.3,2.75);
                \draw [dashed,color=black] (1.3,2.75)--(1.3,-4.7);
                \draw [dashed,color=black] (-1.3,-4.7)--(1.3,-4.7);
                \draw [dashed,color=black] (-1.3,-4.7)--(-1.3,4.1);
                \begin{pgfonlayer}{background layer}
                        \fill[white!10] (-1.3,4.1) rectangle (1.3,-4.7);
                        \fill[white!10] (-1.3,4.1) rectangle (7.3,2.75)node[below,xshift=-5.4cm,yshift=1.0cm]{\color{blue!80!green} \textbf{\large {HEALTHY}}};
                        \draw [dashed,color=black,fill=white!10](4.7,0.9)node[above,yshift=-0.8cm,xshift=4.4cm]{\color{red!80!gray} \textbf{\large{INFECTED}}}rectangle (13.3,-4.7);
                        \draw (1.55,1.8)node[right,xshift=0.3cm,yshift=-3.5cm]{\color{red!70!black}\footnotesize{Immunized}}+(0,-3.5) node[scale=6.0,rotate=0]{\color{gray!40}\}};
                        \draw (7.6,2.5)node[right,xshift=0.3cm,yshift=-4.2cm]{\color{red!70!black} \footnotesize{Positive}}+(0,-4.2) node[scale=6.0,rotate=0]{\color{gray!40}\}};
                \end{pgfonlayer}
     \end{tikzpicture}
                \caption{Population categorization in the CURNDS model. Among these groups, $\boldsymbol{R}$ and $\boldsymbol{C}$ are typically observable, while $\boldsymbol{N}$, $\boldsymbol{U}$, and $\boldsymbol{S}$ represent \underline{unobserved} groups.  The individuals in   $\boldsymbol{D}$    originating from   $\boldsymbol{C}$   can be observed, while those from
$\boldsymbol{U}$   are \underline{unobservable}. }
                \label{Fig_sixgroups}
        \end{figure}
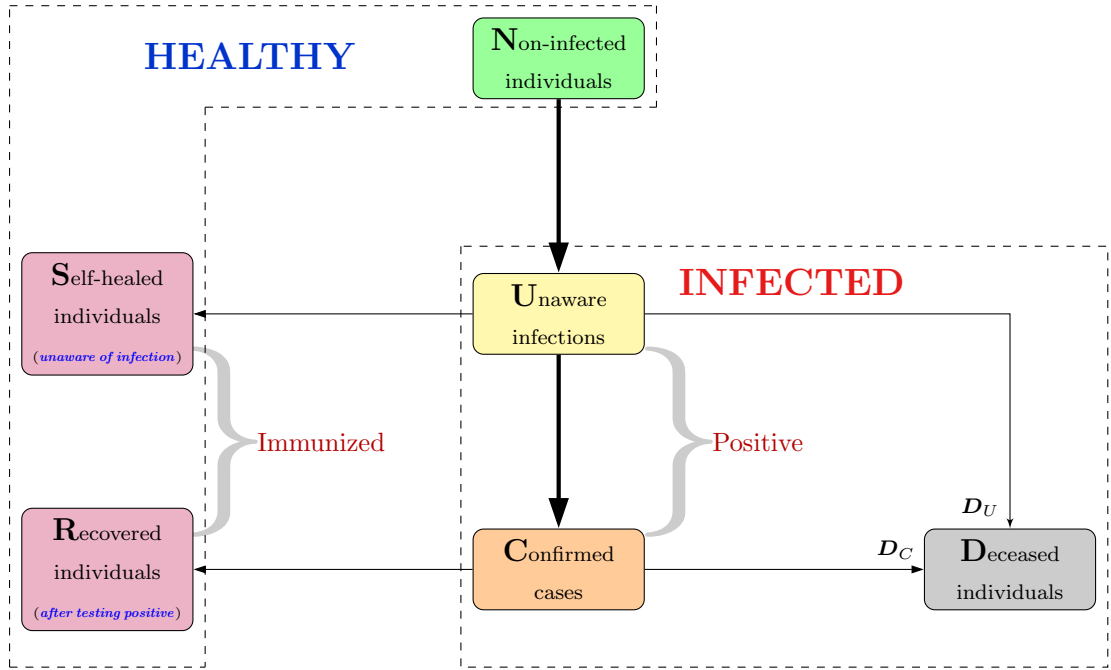

As shown in Figure \ref{Fig_sixgroups}, there are six population groups that fall into two categories (``healthy'' and ``infected'').  Collectively, these groups form the acronym C-U-R-N-D-S.

  The \textbf{healthy} population includes three groups: the \textit{non-infected individuals}  ($\boldsymbol{N}$), the \textit{self-healed individuals unaware of infection} ($\boldsymbol{S}$), and the \textit{recovered individuals after testing positive} ($\boldsymbol{R}$). In detail, $\boldsymbol{N}$ represents the  group of individuals who have not yet been infected, $\boldsymbol{S}$ comprises individuals who, despite being unaware of their infectious status, have managed to heal \emph{without} medical intervention, and $\boldsymbol{R}$ signifies the group of individuals who previously tested positive but have achieved a successful recovery. Notably, we distinguish between the roles of $\boldsymbol{S}$ and $\boldsymbol{R}$ in the transmission dynamics due to the implementation of isolation and other measures associated with the management of quarantinable contiguous diseases.

The \textbf{infected} population can also be divided to three groups: the \textit{unaware infections} ($\boldsymbol{U}$), the \textit{confirmed cases} ($\boldsymbol{C}$), and the \textit{deceased individuals} ($\boldsymbol{D}$). Concretely, $\boldsymbol{U}$ indicates the group of individuals who have contracted the virus but are unaware of their infection, $\boldsymbol{C}$ refers to those who   test  positive for the virus via diagnostic testing, and $\boldsymbol{D}$ denotes the group of individuals who have died during the course of the disease.
 Individuals in the $\boldsymbol{U}$ group may experience delayed diagnoses due to mild symptoms during the incubation period or limited availability of testing resources \cite{zheng2021health}.  Unaware of their condition, these individuals nonetheless remain highly contagious.

It is worth noting that the $\boldsymbol{U}$ and $\boldsymbol{C}$ groups  experience markedly different levels of contact  with other groups---for example, the $\boldsymbol{C}$ group typically has quite limited interactions with non-infected individuals, due to stringent isolation protocols in medical facilities or mandatory home quarantine during the COVID-19 pandemic, while the $\boldsymbol{U}$ group is more likely to have regular interactions with others,  facilitating the spread of the virus \cite{yi2022characterizing, keller2022tracking}.

  The inclusion of groups $\boldsymbol{U}$ and $\boldsymbol{S}$ sets it apart from traditional frameworks like SIR and SEIR. It also diverges from the characterization   of the ``asymptomatic population'' in recent COVID-19 studies \cite{yang2020modified,mwalili2020seir}. According to WHO, the primary mechanism for transmission is \emph{close contact} between individuals, regardless of whether they exhibit symptoms. Indeed,  symptomatic individuals who test positive are usually quarantined effectively to mitigate transmission, while  asymptomatic individuals who remain undiagnosed can continue to spread the virus inadvertently.

Therefore, we advocate  categorizing individuals based on their \textbf{level of contact}, rather than solely on symptom severity or hospitalization status, to have a  more practical understanding of transmission dynamics. In this regard, the CURNDS model provides a more comprehensive yet concise framework than existing literature \citep{yi2022characterizing,mwalili2020seir,keller2022tracking} to better address the challenges posed by highly contagious diseases that require quarantine measures.

   \subsection{Transmissions}\label{Sec_Transmission}
   In this part, we will describe the transmission process associated with CURNDS, which comprises six main links (four of which involve the essential $\boldsymbol{U}$ group). As depicted in Figure \ref{Fig_sixgroups}, most links are self-evident and will be mathematically formulated in the next section. 
 Below we only focus on some critical links, e.g., $\boldsymbol{N}\rightarrow\boldsymbol{U}\rightarrow\boldsymbol{C}$, $\boldsymbol{U}\rightarrow\boldsymbol{S}$ and $\boldsymbol{U}\rightarrow\boldsymbol{D}$, to provide more clarification.

   $\boldsymbol{N}\rightarrow\boldsymbol{U}\rightarrow\boldsymbol{C}$: The transmission from the non-infected group $\boldsymbol{N}$ to the confirmed group $\boldsymbol{C}$ occurs \textbf{exclusively} through individuals in the unaware infected group $\boldsymbol{U}$ (cf. \eqref{EquU}, \eqref{EquC} below). Because our definition of $\boldsymbol{U}$ does not necessarily align with {asymptomatic patients}, its associated transmission dynamics also differ from those in \cite{biswas2020modelling, yi2022characterizing}. Notably,  during the pandemic, isolation measures \cite{chen2020covid} are designed to curtail the virus's spread from confirmed cases, \textit{regardless} of symptom severity, to non-infected individuals.  Moreover, as reported by WHO, many infected individuals were initially unaware of their infection but were highly contagious, often experiencing significant delays in confirmation. The transmission links between  $\boldsymbol{N}$, $\boldsymbol{U}$, and $\boldsymbol{C}$ (see Figure \ref{Fig_N} and the second link in Figure \ref{Fig_U}), though unique, are reasonable representations for quarantinable, highly infectious diseases.

   $\boldsymbol{U}\rightarrow\boldsymbol{S}$ \& $\boldsymbol{U}\rightarrow\boldsymbol{D}$: Infected individuals may self-heal or die before receiving an official diagnosis. These transmissions appear novel compared with the literature \cite{yi2022characterizing,keller2022tracking} and play a vital role in understanding the progression of the disease.

   In addition, it should be noted that the transmission $\boldsymbol{C}\rightarrow\boldsymbol{R}$ is distinct from $\boldsymbol{U}\rightarrow\boldsymbol{S}$,    leading to different transmission rates in the model (cf. \eqref{EquS}, \eqref{EquR} below). An overview of  the transmission process among these groups is shown in Figure \ref{Fig_NUC}. Based on this information, a mathematical model will be used to characterize the evolution of the disease.
   \section{Model Formulation}\label{Sec_Model formulation}
   In this section, we will build a mathematical model based on the groups and transmissions of CURNDS.
   \subsection{A Basic Mathematical Model}\label{Sec_basic math model}
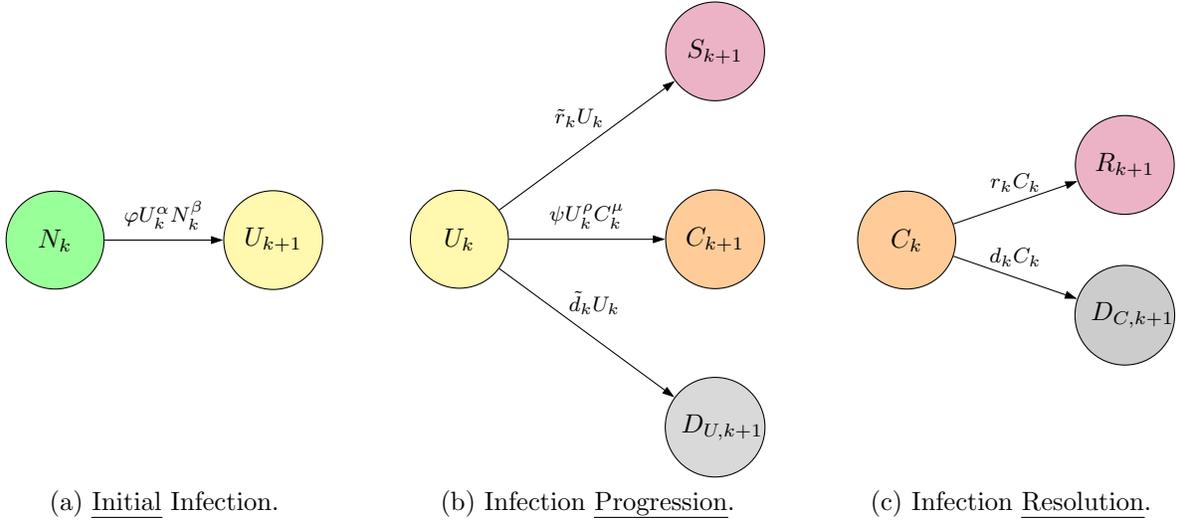
\begin{figure}[ht]
         \centering
        \tikzstyle{groupcircle} = [circle, minimum width = 0.8cm, minimum height=0.5cm, text centered, text width=0.9cm,draw = black]
        \tikzstyle{groupcircleD} = [circle, text width=0.9cm,draw = black]
        \tikzstyle{arrowN} = [arrow,color=green!60]
        \tikzstyle{arrowU} = [arrow,color=yellow!100]
        \tikzstyle{arrowC} = [arrow,color=orange!100]
        \subfloat[\uline{Initial} Infection.\label{Fig_NUCa}]{
                \begin{minipage}[b]{0.32\linewidth}
                        \centering
                        \begin{tikzpicture}[node distance=1.2cm,auto,>=latex',thin]
                                \node[groupcircle, draw = black, fill = green!40](N){\footnotesize $N_k$};
                                \node[groupcircle, right of = N, xshift = 1.7cm,draw = black, fill =yellow!40](U){\footnotesize $U_{k+1}$};
                                \node[groupcircle, below of = N, yshift = -1.35cm,draw = white, fill =white!100](empty1){};
                                \draw [arrow,thin] (N) --node[above]{\scriptsize $\varphi U_k^\alpha N_k^\beta$} (U) ;
                        \end{tikzpicture}
                        \label{Fig_N}
                \end{minipage}%
        }%
        \hfill
        \subfloat[Infection \uline{Progression}.\label{Fig_NUCb}]{
                \begin{minipage}[t]{0.32\linewidth}
                        \centering
                        \begin{tikzpicture}[node distance=1.7cm,auto,>=latex',thin]
                                \node[groupcircle, node distance=1cm,draw = black, fill = yellow!40](U){\footnotesize $U_k$};
                                \node[groupcircle, right of = U, xshift = 1.7cm,draw = black, fill =orange!40](C){\footnotesize $C_{k+1}$};
                                \node[groupcircle, above of = C, yshift = 0.8cm,draw = black, fill =purple!30](S){\footnotesize $S_{k+1}$};
                                \node[groupcircleD, below of = C, yshift = -0.8cm,draw = black, fill =gray!30](DU){\footnotesize $D_{U,k+1}$};
                                \draw [arrow,thin] (U) --node[above,xshift=-0.1cm,yshift=0.1cm]{\scriptsize $\tilde{r}_kU_k$} (S);
                                \draw [arrow,thin] (U) --node[above]{\scriptsize $\psi U_k^\rho C_k^\mu$} (C);
                                \draw [arrow,thin] (U) --node[above,xshift=0.1cm,yshift=0.1cm]{\scriptsize $\tilde{d}_kU_k$} (DU);
                        \end{tikzpicture}
                        \label{Fig_U}
                \end{minipage}%
        }%
        \hfill
        \subfloat[Infection \uline{Resolution}.]{
                \begin{minipage}[t]{0.32\linewidth}
                        \centering
                        \begin{tikzpicture}[node distance=1.2cm,auto,>=latex',thin]
                                \node[groupcircle, node distance=1cm,draw = black, fill = orange!40](C){\footnotesize $C_k$};
                                \node[groupcircle, right of = C, xshift = 1.7cm,yshift = 1cm,draw = black, fill =purple!30](R){\footnotesize $R_{k+1}$};
                                \node[groupcircleD, right of = C, xshift = 1.7cm,yshift = -1cm,draw = black, fill =gray!40](DC){\footnotesize $D_{C,k+1}$};
                                \node[groupcircle, below of = C, yshift = -1.35cm,draw = white, fill =white!100](empty3){};
                                \draw [arrow,thin] (C) --node[above]{\scriptsize $r_kC_k$} (R);
                                \draw [arrow,thin] (C) --node[above]{\scriptsize $d_kC_k$} (DC);
                        \end{tikzpicture}
                \end{minipage}%
        }%
        \caption{Transmission diagrams for CURNDS from time $k$ to $k+1$.}
        \label{Fig_NUC}
     \end{figure}

    Recall that the total population is divided into the $\boldsymbol{N}$, $\boldsymbol{S}$, $\boldsymbol{R}$, $\boldsymbol{U}$, $\boldsymbol{C}$ and $\boldsymbol{D}$ groups. We denote the numbers of individuals in each group at time $k$ by $N_k$, $S_k$, $R_k$, $U_k$, $C_k$, and $D_k$, respectively. For ease of description, the deceased group is further divided into two subgroups: $\boldsymbol{D}=\boldsymbol{D}_{U} \cup \boldsymbol{D}_{C}$, where $\boldsymbol{D}_{U}$ represents the individuals who transitioned from the $\boldsymbol{U}$ group (who remained undiagnosed or untested), while $\boldsymbol{D}_{C}$ comprises the individuals who had  been previously confirmed as positive in the $\boldsymbol{C}$ group. In this way, the model can be decomposed into three components: a) \textit{initial infection},  b) \textit{infection progression}, and c) \textit{infection resolution}, as illustrated in Figure \ref{Fig_NUC}.

     \paragraph{Initial Infection}
     Traditional epidemiological models often assume that the number of newly added infections at time $k$, represented by $\Delta N_k$, is proportional to $N_k$ and $U_k$, i.e., $$\Delta N_k \propto - U_k N_k.$$ This  relies on the concept of \textit{uniform mixing}, where infected and non-infected individuals interact randomly and homogeneously within the population \cite{anderson1991}. However, from our extensive experience, $\Delta N_k$ may grow \textbf{sublinearly} with respect to either $N_k$ or $U_k$; sometimes, $U_k$ may dominate the influence, while $N_k$'s effect is characterized by a much smaller power index (refer to the section on contrasting power parameters for more details). 
 Therefore, we assume
 \begin{equation}\label{EquN}
        N_{k+1}=N_k-\varphi U_k^\alpha N_k^\beta,
 \end{equation}
  where $\varphi$ is a rate parameter for infection, and $\alpha$ and $\beta$ are the power parameters to be estimated, measuring the impact of $U_k$ and $N_k$, respectively. Standard models adopt $\alpha=\beta=1$. Introducing such a nonlinear dependence on $U_k$ or $N_k$ is new but effective in the estimation of epidemiological models (see the section on forecasting for further details). 

  \paragraph{Infection Progression} Because some individuals in the $\boldsymbol{U}$ group may self-heal and transition to the $\boldsymbol{S}$ group, we propose
    \begin{equation}\label{EquS}
        S_{k+1}=S_k+\tilde{r}_k U_k.
  \end{equation}
   Similarly, we assume
    \begin{equation}\label{EquDU}
        D_{U,k+1}=D_{U,k}+\tilde{d}_k U_k.
   \end{equation}
   Here, we allow the \textit{self-healing rates} $\tilde{r}_k$ and the \textit{death rates of the unaware infected population} $\tilde{d}_k$ to \textbf{vary over time}, reflecting the evolving nature of the virus and adapting healthcare responses \cite{gong2023sars}.  Compared with the classical SEIR model, we have found that a constant rate parameter works poorly for modeling persistent diseases, such as COVID-19, probably due to a variety of heterogeneities caused by the prolonged duration or isolation measures. The large number of unknown parameters resulting from the introduction of time varying rates will be addressed using a {functional} approach in the next subsection. 

    Of course, the individuals in the $\boldsymbol{U}$ group may also test positive and transition to the $\boldsymbol{C}$ group. Similar to Equation \eqref{EquN}, we can introduce a rate parameter $\psi$ and two power indices $\rho$ and $\mu$ to quantify the effect of $U_k$ and $C_k$ on the number of newly confirmed cases (cf. Equation \eqref{EquU}). This is because rises in  $U_k$   and  $C_k$ drive up diagnoses and broaden public testing, respectively.     \paragraph{Infection Resolution} The individuals in the $\boldsymbol{C}$ group may move to the $\boldsymbol{R}$ group or the $\boldsymbol{D}_{C}$ group. Based on similar arguments, we adopt the following model,
        \begin{numcases}{}
        R_{k+1}=R_k+r_k C_k, \label{EquR}\\ 
        D_{C,k+1}=D_{C,k}+d_k C_k,\label{EquDC} 
   \end{numcases}
   where $r_k$ and $d_k$ are the {time varying} \textit{recovery rates} and \textit{death rates} for the confirmed group, respectively. Correspondingly, $U_{k+1}$ and $C_{k+1}$ are given by
        \begin{numcases}{}
        U_{k+1}=U_k+\varphi U_k^\alpha N_k^\beta-\psi U_k^\rho C_k^\mu-\tilde{r}_k U_k-\tilde{d}_k U_k\label{EquU},\\ 
        C_{k+1}=C_k+\psi U_k^\rho C_k^\mu-r_k C_k-d_k C_k.\label{EquC}
    \end{numcases}
   We emphasize the importance of distinguishing between the rate parameters ($\tilde{r}_k$ and $\tilde{d}_k$) for the $\boldsymbol{U}$ group from those ($r_k$ and $d_k$) for the $\boldsymbol{C}$ group. This is because, for instance, inadequate medical attention can lead to reduced self-recovery rates and increased mortality among unaware infected individuals compared to treated cases.   

\subsection{Statistical Modeling and Smoothing }\label{Sec_chanllenge}
   First, among the seven sequences defined previously, 
 $R_k$, $C_k$, and $D_{C,k}$, which appear at the bottom level of Figure \ref{Fig_sixgroups}, are typically observed but may contain noise. Therefore, to deal with real-world data, Equations \eqref{EquR}, \eqref{EquDC} and \eqref{EquC} should be rephrased as:
        \begin{numcases}{}
                R_{k+1}=R_k+r_k C_k+\varepsilon_{R,k+1},\label{EquRerror}\\
                D_{C,k+1}=D_{C,k}+d_k C_k+\varepsilon_{D,k+1},\label{EquDCerror}\\
                C_{k+1}=C_k+\psi U_k^\rho C_k^\mu-r_k C_k-d_k C_k+\varepsilon_{C,k+1}, \label{EquCerror}
        \end{numcases}
   where $\varepsilon_{R,k+1}, \varepsilon_{D,k+1}, \varepsilon_{C,k+1}$ denote the random errors at time $k+1$ ($0\le k \le n-1$).

   The challenge in statistical modeling arises from the limited number of observations ($3n+3$) in comparison to the large number of free parameters ($4n+10$), leading to overfitting issues \cite{hawkins2004problem}. While some additional information can be used, this issue is mainly caused by time-varying transmission rates, $r_k$, $d_k$, $\tilde{r}_k$ and $\tilde{d}_k$. We model these sequences as discrete-time functions, leveraging the fact that regional medical conditions typically give rise to \textit{smoothly} varying rate functions over time \cite{lai2020effect}. In light of this, we propose using a smooth nonparametric modeling approach for each time varying rate sequence, which can reduce the number of unknown parameters and result in a more accurate estimation of the dynamics.

   We take the recovery rates $r_k$ as an example,   now  represented as a function $r(\cdot)$ evaluated at discrete time points  $k=0,1,\cdots,n-1$. When dealing with a smooth nonparametric function, one common approach is to approximate it using piecewise polynomials, or splines \cite{wahba1990spline}. This  involves specifying a limited number of interior knots located at, say, $0<t_1<\cdots <t_j<\cdots <t_{m_r}<n-1$, and requires continuity constraints of the function values and first and second derivatives at the knots, along with specified boundary conditions, to ensure a smooth fit and an accurate approximation using fewer parameters. Here, $m_r$ denotes the number of interior knots for the $r$-sequence. Let $r_j(\cdot)$ be the cubic polynomial on $[t_j, t_{j+1}]_{1\leq j\leq m_r}$, subject to the conditions $r_{j}(t_j)=r_{j+1}(t_j)$, $ r_{j}^{\prime}(t_j)=r_{j+1}^{\prime}(t_j)$, and $ r_{j}^{\prime\prime}(t_j)=r_{j+1}^{\prime\prime}(t_j)$. The natural boundary conditions  $r_{1}^{\prime\prime}(t_1)=0$ and $r_{m_r}^{\prime\prime}(t_{m_r})=0$
    are also imposed to produce stable predictions and reduce overfitting \cite{hastie2009elements}. In this way, the number of parameters for the recovery rates can be reduced from $n$ to $m_r+2$. Empirically, we found 3 to 8 knots suffice. Similar treatments apply to
  $d_k$, $\tilde{r}_k$, and $\tilde{d}_k$.\\

  In addition, some databases often provide data on the number of daily conducted tests \cite{Hasell2020}, which can offer valuable information for estimating model parameters. Because testing is generally sought by individuals who are uncertain about their infection status, we assume the number of tests, $T_{k+1}$, is associated with the predefined sequences by
   \begin{equation}\label{EquTerror}
        T_{k+1}=T_k+h(k)(N_k+S_k+U_k)+\varepsilon_{T,k+1},
   \end{equation}
  where $\varepsilon_{T,k+1}$ represents the random error at time $k+1$, and $h(\cdot)$ is the rate function for $T_{k+1}$.

We also assume that the total population size at the beginning of the epidemic, denoted as $P_0$,  is available, and so
  \begin{equation}\label{EquP}
        N_0+U_0+S_0+D_{U,0}+R_{0}+C_{0}+D_{C,0} = P_0.
  \end{equation}
  In our model, we enforce the total population constraint \emph{only} at the initial time $t=0$. In noise-free scenarios, the sum of the dynamical equations naturally preserves the total population size over time. However, real-world data often contain measurement errors, and for longer-term studies in a specific region,  factors like migration, birth rates and mortality can make enforcing this constraint at \emph{every} time point {counterproductive}  to model accuracy \cite{yang2020modified,Zhan2020Modeling}.

   In summary, with a spline formulation of the rates and the inclusion of Equations \eqref{EquRerror}--\eqref{EquP}, we observe $4n+4$ data points. However, with a fixed   number of knots, the model retains a constant number of free parameters (under 60 in our analysis). 

  \section{Computation}\label{Sec_computation}

        First, we  reparameterize the problem to simplify the computation. The observable sequences $R_{k}$, $C_{k}$, $D_{C, k}$, and $T_{k}$ are denoted as $X_{i, k+1}$ for $1 \leq i \leq 4$, and the sequences $N_{k}$, $S_{k}$, $U_{k}$, and $D_{U, k}$ are denoted as $Z_{j, k+1}$ for $1 \leq j \leq 4$.  The model   in Equations \eqref{EquRerror}--\eqref{EquTerror} can be expressed as
        \begin{subequations}
        \begin{numcases}{}
        X_{1,k+1}=X_{1,k}+r(k)X_{2,k}+\varepsilon_{1, k+1},\label{EquX1}\\      X_{2,k+1}=(1-r(k)-d(k))X_{2,k}+\psi Z_{3,k}^{\rho} X_{2,k}^{\mu}+\varepsilon_{2, k+1},\label{EquX2}\\
        X_{3,k+1}=X_{3,k}+d(k)X_{2,k}+\varepsilon_{3, k+1},\label{EquX3}\\
        X_{4,k+1}=X_{4,k}+h(k)(Z_{1,k}+ Z_{2,k}+ Z_{3,k})+\varepsilon_{4, k+1}.\label{EquX4}
    \end{numcases}
\end{subequations}
The   estimation problem is subject to    the following constraints
\begin{subequations}
\begin{numcases}{}
        Z_{1,k}=Z_{1,k-1}-\varphi Z_{3,k-1}^{\alpha} Z_{1,k-1}^{\beta},\label{EquZ1}\\
        Z_{2,k}=Z_{2,k-1}+\tilde{r}(k-1) Z_{3,k-1},\label{EquZ2}\\
        Z_{3,k}=(1-\tilde{r}(k-1) -\tilde{d}(k-1) )Z_{3,k-1}+\varphi Z_{3,k-1}^{\alpha} Z_{1,k-1}^{\beta} -\psi Z_{3,k-1}^\rho X_{2,k-1}^\mu, \label{EquZ3}\\
        Z_{4,k}=Z_{4,k-1}+\tilde{d}(k-1) Z_{3,k-1},\label{EquZ4}
\end{numcases}
\end{subequations}
   where $\alpha,\beta,\rho,\mu >0$ are the power parameters, $\varphi, \psi \in (0,1)$ are the   rate parameters, the random observation errors, $ \varepsilon_{i, k+1} ( 1\leq i \leq4)$ are independent of  $X_{i,k}$ ($1\leq i \leq4$). One can assume   $X_{i,k}$ to be   Poisson; other  models include $X_{i,k}$ being modeled as negative binomial, or $\varepsilon_{i, k+1}$ as log-normal.

As discussed in the previous section, 
 we advocate the use of natural cubic splines to model the rate functions $r(\cdot)$, $d(\cdot)$, $h(\cdot)$, $\tilde{r}(\cdot)$, and $\tilde{d}(\cdot)$ due to their inherent smoothness. This spline-based representation simplifies the parameterization of these functions.

For instance, the rate function $r(\cdot)$ is represented as: \begin{equation} r(\cdot) = \sum_{j=1}^{m_r+2} \theta_{r,j}V_{r, j}(\cdot), \end{equation} where $m_r$ is the number of knots,    ${\boldsymbol \theta}_r =[ \theta_{r, j}]= [\theta_{r, 1}, \theta_{r, 2}, \cdots, \theta_{r, m_r+2}]^\mathsf{T}$
 represents the vector of basis expansion coefficients, and $V_{r,j}(\cdot)$ denotes the spline basis functions. With the knots selected, the spline functions $V_{r,j}(\cdot)$ are known, but the coefficients ${\boldsymbol \theta}_r$ remain to be determined.

   The functions $d(\cdot)$, $\tilde{r}(\cdot)$, $\tilde{d}(\cdot)$, and $h(\cdot)$ can be modeled using a similar framework. Each is expressed as a linear combination of spline basis functions with their respective coefficient vectors:  ${\boldsymbol \theta}_d$, ${\boldsymbol \theta}_{\tilde{r}}$, ${\boldsymbol \theta}_{\tilde{d}}$, and ${\boldsymbol \theta}_h$. Leveraging the smoothness of these functions not only reduces the number of unknown parameters but also enhances the accuracy of predictions.

After the basis expansion, the computational challenge simplifies to a standard constrained optimization problem. The parameters can be optimized using interior point optimizer, a large-scale nonlinear optimization software library ideal for smooth objectives with constraints. We employed the ipopt  library \cite{Wachter2006} to solve the non-convex optimization problem. Due to the problem's nonconvexity, a multi-stage screening approach with multiple initializations proves effective. We  first generate $K_0$ starting points. The algorithm then runs for $I_0$ iterations, after which it selects the $K_1$ best candidates based on their objective function values. The final estimate is the one among these candidates that yields the lowest objective function value. Scalar parameters such as power indices $\alpha$, $\beta$, $\rho$, and $\mu$, which are bounded between $0$ and $1$, can be initialized using a crude grid search with 3 to 5 values.  Among the parameters, $r_k$ and $d_k$ exhibit higher sensitivity. Fortunately,  preliminary estimates for these parameters can be obtained using merely $X_{1,k}$, $X_{2,k}$, and $X_{3,k}$ (cf. Equations \eqref{EquDC} and \eqref{EquC}) through   spline approximation. Though less accurate than joint optimization estimates, these initial values serve as suitable starting points for subsequent optimization iterations.

\section{Case Study: Forecasting with Quebec Covid-19 Data}\label{Sec_Real dataset analysis}

We have conducted extensive forecasting experiments with the CURNDS model and other established methods \cite{alzahrani2020forecasting, yonar2020modeling, Guleryuz2021forecasting}. In this section, we present a practical case study using COVID-19 data from Quebec, Canada \cite{canadagov}. All experiments were performed using Python 3.7.10 on an Intel Core I7-8700 processor with 32GB RAM.
The dataset comprises daily records of recovered individuals, confirmed cases, deceased individuals, and test data. Consistent with standard practices in the field \cite{saraiva2023hierarchical}, we processed the raw data using a 7-day moving average to smooth out fluctuations and reduce noise.

Below, we compare CURNDS with SEIR, ARIMA, Holt's Exponential Smoothing (ES), and Long Short-Term Memory (LSTM) model. SEIR, a conventional framework for modeling infectious disease dynamics, is widely used in epidemic progression analysis. ARIMA, a popular time series method, is often effective for sequential data forecasting \cite{alzahrani2020forecasting}. The ES approach we utilize, as detailed in \citep{holt2004forecasting}, effectively captures seasonal and trend components within data; similar methods have demonstrated effectiveness in handling dynamic data patterns  \cite{yonar2020modeling}. LSTM, a neural network model, has  gained considerable success across various applications. Each of these models has been chosen for its specific strengths, contributing to a comprehensive forecasting comparison \cite{Guleryuz2021forecasting}.

\subsection{5-day Progressive Forecast}

This part carries out a progressive forecasting experiment, where each method is trained on a dataset from September 1st to 29th, 2020---a period marked by a high number of newly confirmed cases. We then conduct a 5-day forecast, during which each prediction is based on the estimated parameters and the preceding forecasts.


Evaluation metrics include an absolute error, denoted by $\text{Err}_{\text{\scriptsize abs}} = \lvert Y_k - \widehat{Y}_k \rvert$, which is robust in measuring discrepancies in cases where the data distribution is typically non-normal. We also calculate  a scale-independent relative error, or absolute percentage error, represented by $\text{Err}_{\text{\scriptsize rel}} = \frac{\lvert Y_k - \widehat{Y}_k \rvert}{Y_k} \times 100\%$.
Mean and median values of these errors for the observable sequences of $R_k$, $C_k$, and $D_{C, k}$   are reported in Table \ref{pred_errorOR_5day}. 

 \begin{table}[hbtp]
        \renewcommand{\arraystretch}{1.3}
        \setlength{\tabcolsep}{1.2mm}{
                \begin{tabular}{rrcccccccccccccc}
                        \toprule[1pt]
                        \hline
                        &&\multicolumn{4}{c}{\textbf{\small Recovered} ($R_k$)}&&\multicolumn{4}{c}{\textbf{\small Confirmed} ($C_k$)} &&\multicolumn{4}{c}{\textbf{\small Deceased} ($D_{C,k}$)}\\        \cmidrule(r){3-6}\cmidrule(r){8-11}\cmidrule(r){13-16}
                        && \multicolumn{2}{c}{$\text{Err}_\text{\scriptsize abs}$} & \multicolumn{2}{c}{$\text{Err}_\text{\scriptsize rel}$} &&   \multicolumn{2}{c}{$\text{Err}_\text{\scriptsize abs}$} & \multicolumn{2}{c}{$\text{Err}_\text{\scriptsize rel}$}&&   \multicolumn{2}{c}{$\text{Err}_\text{\scriptsize abs}$} & \multicolumn{2}{c}{$\text{Err}_\text{\scriptsize rel}$}\\        \cmidrule(r){3-4}\cmidrule(r){5-6}\cmidrule(r){8-9}\cmidrule(r){10-11}\cmidrule(r){13-14}\cmidrule(r){15-16}
                        &&mean&med&mean& med&&mean&med&mean& med&&mean&med&mean& med \\ \hline
                        \textbf{CURNDS} &&10.0&   10.9  & 1.58 & 1.74&  &18.1 &8.05& 28.7& 13.6 &&1.77 & 1.92&3.02& 3.29\\
                        \textbf{SEIR} &&264&   237 & 41.8 & 38.0&& 403 & 433&677& 670& & 2.50 &1.81&4.27& 3.09\\
                        \textbf{ARIMA} &&90.5& 64.4&14.2 &  9.40&& 91.7& 91.1 &147 & 113&& 8.54& 2.68 & 14.6& 13.2 \\
                        \textbf{ES} &&355& 299 & 56.1&  47.8  && 130 & 87.6&207 & 147 && 8.32 & 7.47& 14.2 & 12.8\\
                        \textbf{LSTM} &&87.4& 88.9 &13.9&  14.2&&391 & 355& 642& 599 && 2.74 & 3.20& 4.68& 5.48   \\
                        \bottomrule[1pt]
        \end{tabular}}
        \centering
        \caption{Comparison of progressive forecast errors for the recovered individuals ($R_k$), the confirmed cases ($C_k$), and the deceased individuals from the confirmed group ($D_{C,k}$), in terms of mean and median absolute and relative errors. }
        \label{pred_errorOR_5day}
\end{table}

Table \ref{pred_errorOR_5day} presents a forecast error comparison between different methods. The SEIR method, widely-used  for modeling infectious diseases, demonstrates suboptimal performance, particularly in predicting $C_k$, and its mean absolute error exceeds that of the CURNDS model by more than 24 times.   ARIMA and ES models   deliver unsatisfactory results in forecasting $D_{C,k}$, with mean absolute errors at least three times higher than those of other methods. ES exhibits poor performance in predicting $R_k$.   LSTM shows mixed results: reasonable accuracy   for $R_k$ and $D_{C,k}$, but poor performance for $C_k$. Overall, the CURNDS model outperforms its counterparts, consistently delivering the lowest error rates across various scenarios.

\begin{figure}[t]
        \centering
        \subfloat[{\scriptsize No. of Recovered}]{
                \begin{minipage}[ht]{0.33 \linewidth}
                        \centering
                        \includegraphics[scale=0.27]{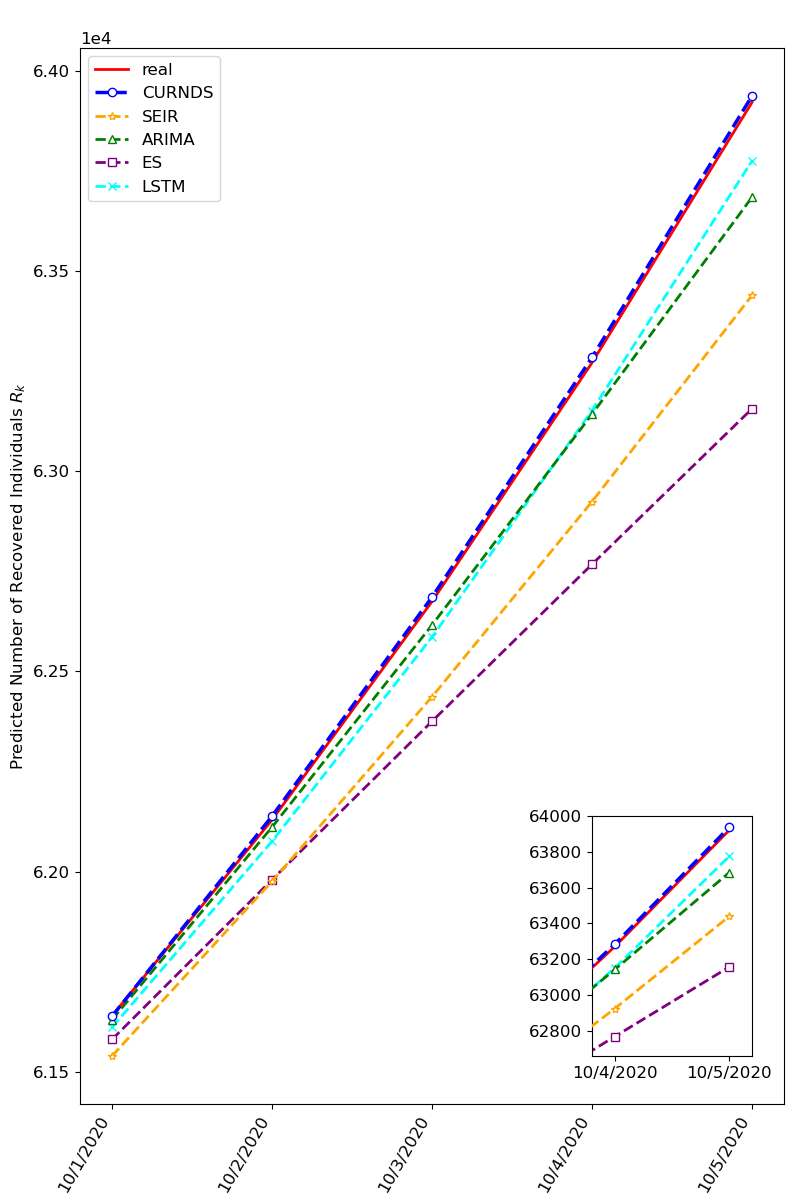}
                \end{minipage}%
        }%
        \centering
        \subfloat[{\scriptsize No. of Confirmed}]{
                \begin{minipage}[ht]{0.33\linewidth}
                        \centering
                        \includegraphics[scale=0.27]{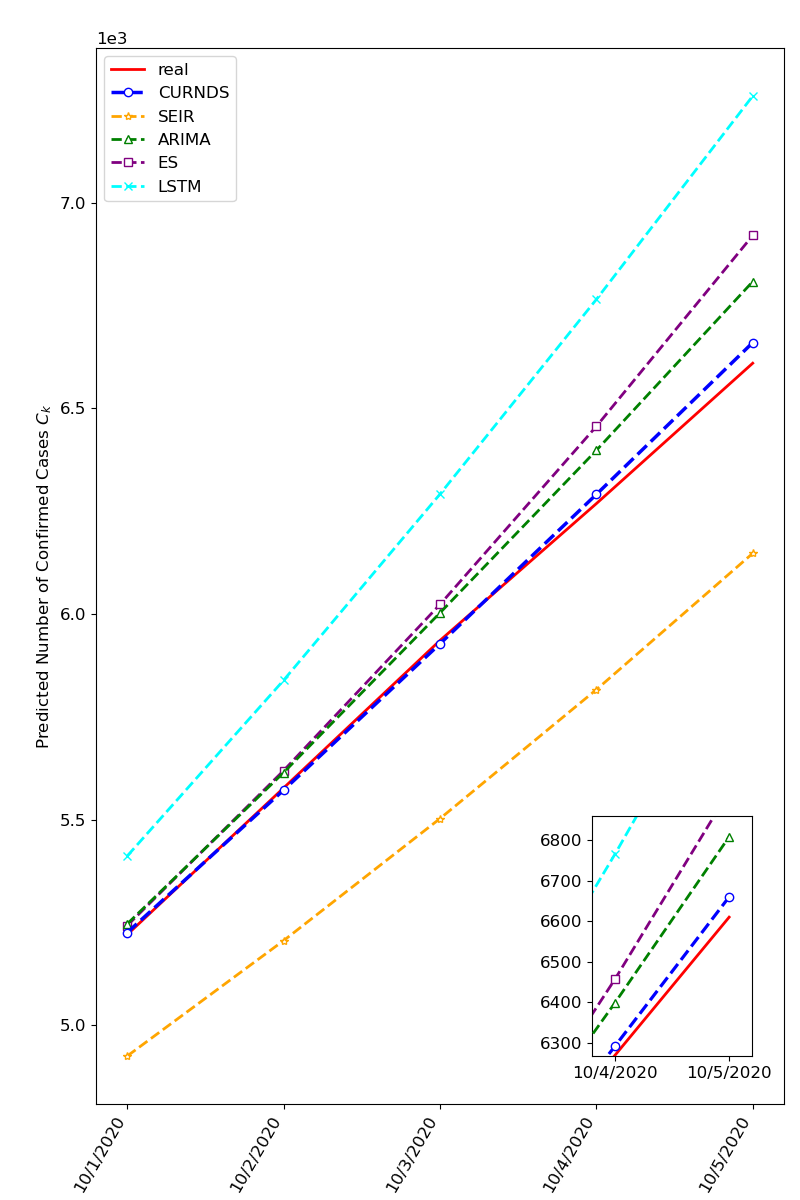}
                \end{minipage}%
        \label{fig_progress_fore_confirm}
        }%
        \centering
        \subfloat[{\scriptsize No. of Deceased from Confirmed Cases}]{
                \begin{minipage}[ht]{0.33\linewidth}
                        \centering
                        \includegraphics[scale=0.27]{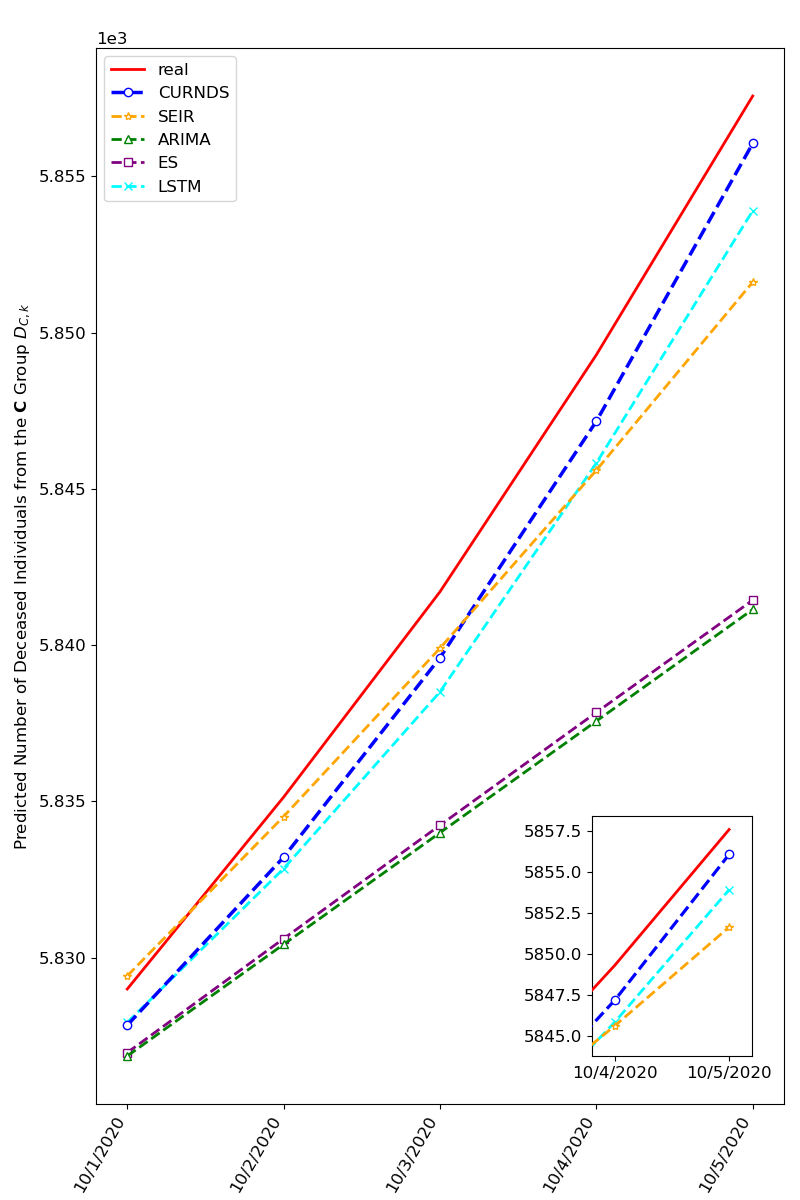}
                \end{minipage}%
        \label{fig_progress_fore_death}
        }%
        \centering
        \caption{A visual comparison of progressive forecast performances for the number of recovered individuals ($R_k$), the confirmed cases ($C_k$), and the deceased individuals from the confirmed group ($D_{C,k}$). As time progresses in the prediction process, the cumulative errors tend to increase for most models. Our model (blue) consistently maintains a close alignment with the real values (red). The y-axis \emph{scale} and \emph{range} have been adjusted  in each plot to ensure clear visualization of all curves.}
        \label{fig_5-day-fore}
\end{figure}

Figure \ref{fig_5-day-fore}   illustrates the forecasting results. As time advances, the accuracy of all models degrades, particularly for  $C_k$ and $D_{C,k}$ predictions. However, the curves of the CURNDS model (depicted in blue)   remain closer to the actual values (depicted in red), indicating the model's robust performance in maintaining accuracy in progressive forecasts.


\subsection{Rolling Forecast}
Next, we performed a \textit{rolling} forecasting experiment to evaluate the dynamic performance of the CURNDS model against other methods. This approach involves continuously updating the training dataset with the most recent data, which enables the model to adjust to rapidly changing conditions.

Specifically, with the observed data series ${Y_1, Y_2, \cdots, Y_k, \cdots, Y_n}$ and a predefined window size $w$ (where $w < n$), the window shifts forward by one time step after each forecast, integrating the latest data. At any given time $k$, the model uses the current window of data, ${Y_{k-w+1}, Y_{k-w+2}, \ldots, Y_k}$, to predict the following day’s value, $\widehat{Y}_{k+1}$. The experiment begins with training data starting on September 1st, 2020, using a window size of 30 days.

\begin{table}[h]
        \renewcommand{\arraystretch}{1.3}
        \setlength{\tabcolsep}{1.2mm}{
                \begin{tabular}{rrcccccccccccccc}
                        \toprule[1pt]
                        \hline
                        &&\multicolumn{4}{c}{\textbf{\small Recovered} ($R_k$)}&&\multicolumn{4}{c}{\textbf{\small Confirmed} ($C_k$)} &&\multicolumn{4}{c}{\textbf{\small Deceased} ($D_{C,k}$)}\\        \cmidrule(r){3-6}\cmidrule(r){8-11}\cmidrule(r){13-16}
                        && \multicolumn{2}{c}{$\text{Err}_\text{\scriptsize abs}$} & \multicolumn{2}{c}{$\text{Err}_\text{\scriptsize rel}$} &&   \multicolumn{2}{c}{$\text{Err}_\text{\scriptsize abs}$} & \multicolumn{2}{c}{$\text{Err}_\text{\scriptsize rel}$}&&   \multicolumn{2}{c}{$\text{Err}_\text{\scriptsize abs}$} & \multicolumn{2}{c}{$\text{Err}_\text{\scriptsize rel}$}\\        \cmidrule(r){3-4}\cmidrule(r){5-6}\cmidrule(r){8-9}\cmidrule(r){10-11}\cmidrule(r){13-14}\cmidrule(r){15-16}
                        &&mean&med&mean& med&&mean&med&mean& med&&mean&med&mean& med \\ \hline
                        \textbf{CURNDS} &&12.2 &  8.87& 1.88& 1.33 & &  35.3 &14.2& 46.2& 19.2 && 1.19&  1.08 &2.03 & 1.84\\
                        \textbf{SEIR} & & 163& 146 &25.0 & 22.8 && 293 &  247& 420& 426&&7.24&  8.89 &12.3& 15.2 \\
                        \textbf{ARIMA} & & 14.2& 12.2 & 2.20& 1.96 & &25.7 & 22.5& 38.2& 39.2& &1.06& 0.93&1.80 & 1.59\\
                        \textbf{ES} && 60.3  & 59.3& 9.35& 9.35 && 36.6 & 35.1& 51.5& 49.3&&1.20& 1.14 & 2.04& 1.94 \\
                        \textbf{LSTM} & & 84.8 & 94.7& 13.1& 14.4& & 274 & 266&408& 419&&2.05&  2.03&3.50 & 3.46 \\
                        \bottomrule[1pt]
        \end{tabular}}
        \centering
        \caption{Comparison of rolling forecast errors for the recovered individuals ($R_k$), the confirmed cases ($C_k$), and the deceased individuals from the confirmed group ($D_{C,k}$).}
        \label{rolling_pred_error}
\end{table}

Table \ref{rolling_pred_error} presents the rolling forecast errors for various models. Echoing the findings from Table \ref{pred_errorOR_5day}, the SEIR model continues to exhibit inferior performance across all scenarios in this experiment. While LSTM's errors are marginally lower than those of the SEIR model, they remain substantially higher than those obtained by the other models. In comparison with the   progressive forecasts, ARIMA and ES demonstrate enhanced accuracy in estimating $C_k$.  CURNDS consistently outperforms other methods, ranking either first or second in all evaluated scenarios.



\section{Characteristics Study: Wild-Type Strain vs. Omicron Variant}\label{Characteristics Study}
Viruses, including COVID-19, naturally mutate, giving rise to new variants with different disease characteristics. This section undertakes a study to examine the transmission dynamics between the ``wild-type strain" and the ``Omicron variant." We selected two specific periods of data from Quebec: (a) the  \textbf{wild-type period}  from September 1st to 29th, 2020, during which the wild-type strain of COVID-19 accounted for over 99\% of daily new cases, and (b) the  \textbf{Omicron-variant period} from February 9th to March 9th, 2022. Using the CURNDS model, we aim to estimate dynamic factors related to both strains, to better understand the virus's evolving behavior and its implications for public health and preventive measures.

\subsection{Contrasting  {Power} Parameters}
\label{subsec:powercontrast}
Table \ref{univariate_paramaters} presents the estimated power indices for both the wild-type and Omicron-variant periods.

       \begin{table}[h]
        \renewcommand\arraystretch {1.2}
        \setlength{\tabcolsep}{5pt}
        \centering
        \begin{tabular}{clcc}
                \toprule[1pt]
                \hline
                &  &\textbf{\small Wild-type} & \textbf{\small Omicron-variant}\\
                \hline
                 \multirow{2}{*}{\footnotesize Infection Power Indices}&$\hat{\alpha}$ & 0.9& 0.6\\
                &$\hat{\beta}$ & 0.05 & 0.2\\ \hline
                \multirow{2}{*}{\footnotesize Confirmation Power Indices}
                &$\hat{\rho}$& 0.3 & 0.5 \\
                &$\hat{\mu}$ & 0.4 & 0.4 \\
                \bottomrule[1pt]
        \end{tabular}
        \caption{The estimated power parameters in the infection and confirmation processes for both the wild-type and the Omicron-variant. }
        \label{univariate_paramaters}
\end{table}

We begin by analyzing the estimates of $\alpha$ and $\beta$ within each period, reflecting the impact of unaware infections ($U_k$) and non-infected individuals ($N_k$) on new infections, respectively (cf. Figure \ref{Fig_NUCa}). 
  Conventionally, power indices are fixed at 1, presupposing homogeneous mixing between infected and susceptible populations \cite{anderson1991}. However, government-imposed isolation and public health interventions likely violate the assumption for COVID-19 \cite{chen2020covid}.   Our analysis shows \emph{sublinear} infection growth relative to  $ U_k$ or $N_k$, with power indices significantly below 1 (except for the power of $U_k$ during the initial wild-type period), reflecting reduced contact rates due to mask-wearing, social distancing, and other risk-reduction strategies \cite{howard2021evidence}. 

 Furthermore, the substantial disparity between   $\widehat{\alpha}$ and $\widehat{\beta}$     underscores distinct roles of unaware infections     ($U_k$) and non-infected individuals ($N_k$) to new cases. By adaptively estimating power parameters, CURNDS captures non-uniform interaction patterns, improving model  applicability.

Our analysis of confirmation power indices   $\rho$ and $\mu$,  which quantify  the impact of unaware infections ($U_k$) and confirmed cases ($C_k$) on new confirmations (cf. Figure \ref{Fig_NUCb}),  consistently approximate  0.5 for both the wild-type and Omicron-variant periods, significantly  below 1.  This indicates a sublinear relationship, where newly confirmed cases grow proportionally to the square root of $U_k$ and $C_k$,  rather than linearly.

\subsection{Transmission Dynamics}
\begin{figure}[h]
        \centering
        \includegraphics[scale=0.4]{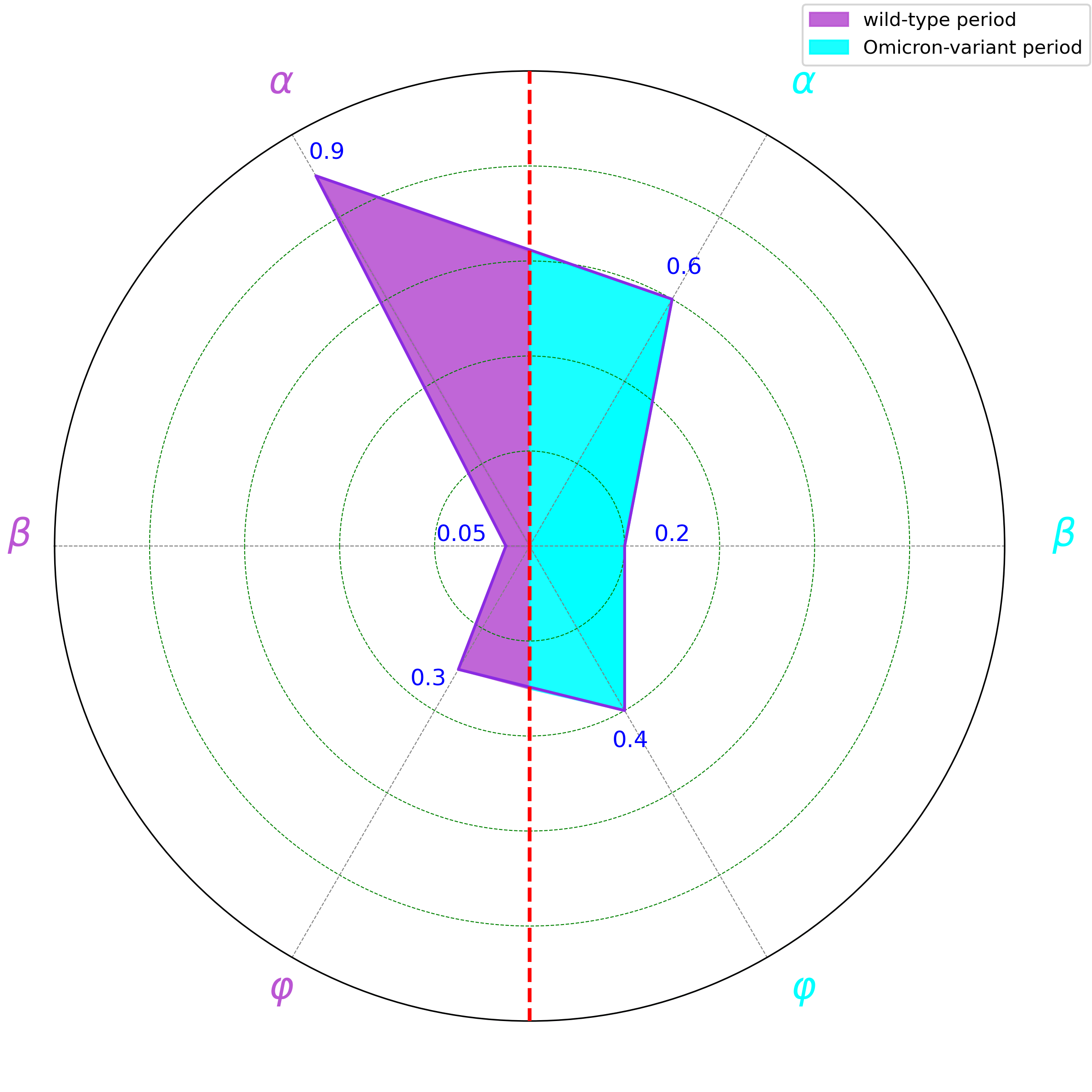}
        \centering
        \caption{Comparison of infection parameters $\alpha$, $\beta$, $\varphi$ between {wild-type} and {Omicron-variant} periods.}
        \label{fig_infe}
\end{figure}

Next, let us study the transmission dynamics of virus across the different periods. The process involves \textit{infection} parameters like rate $\varphi$, power indices $\alpha$, $\beta$, \textit{confirmation} parameters including rate $\psi$, power indices $\rho$, $\mu$, as well as the {time-varying} sequences $\tilde{r}_k$ and $\tilde{d}_k$.

\begin{figure}[h]
        \centering
        \includegraphics[scale=0.4]{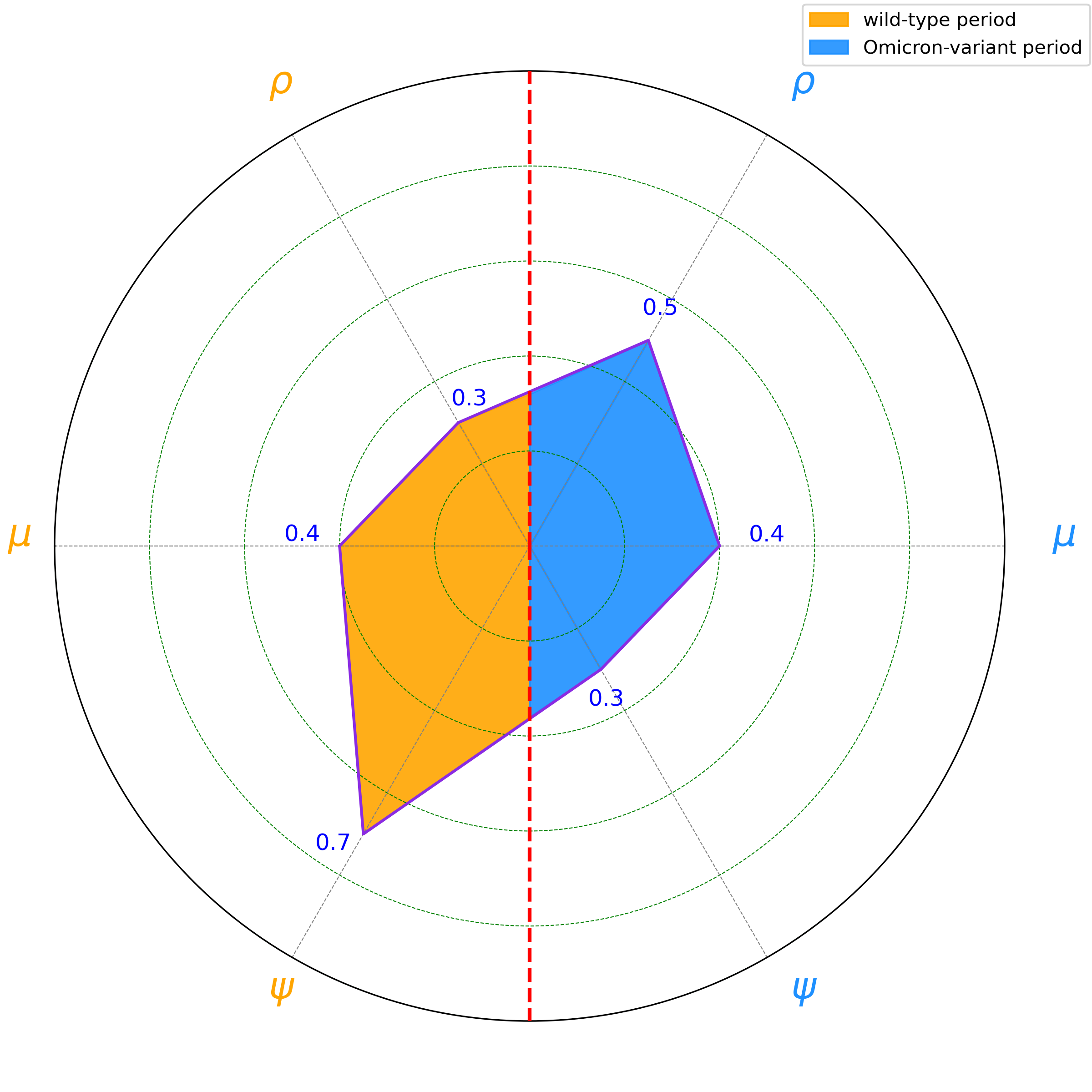}
        \centering
        \caption{Comparison of confirmation parameters $\rho$, $\mu$, $\psi$, between {wild-type} and {Omicron-variant} periods.}
        \label{fig_conf}
\end{figure}

\begin{figure}[htbp]
        \centering
        \subfloat[{\scriptsize Time varying self-healing rates $\tilde{r}_k$}.]{
                \begin{minipage}[h]{0.5\linewidth}
                        \centering
                        \includegraphics[scale=0.43]{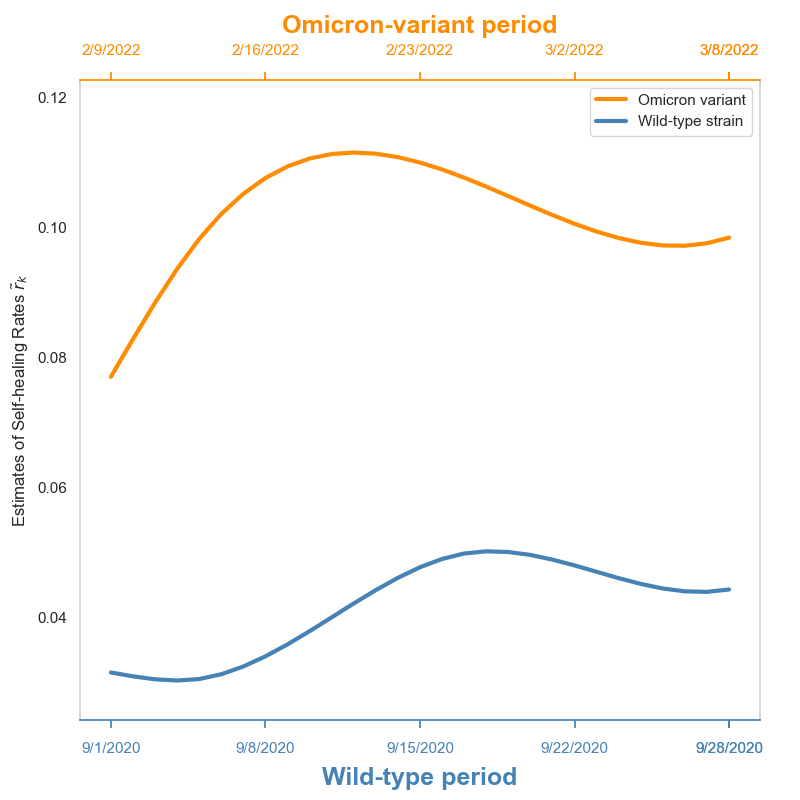}
                        \label{fig_rTld_line}
                \end{minipage}%
        }%
        \subfloat[{\scriptsize Violin plot of the self-healing rates.}]{
                \begin{minipage}[h]{0.5\linewidth}
                        \centering
                        \includegraphics[scale=0.43]{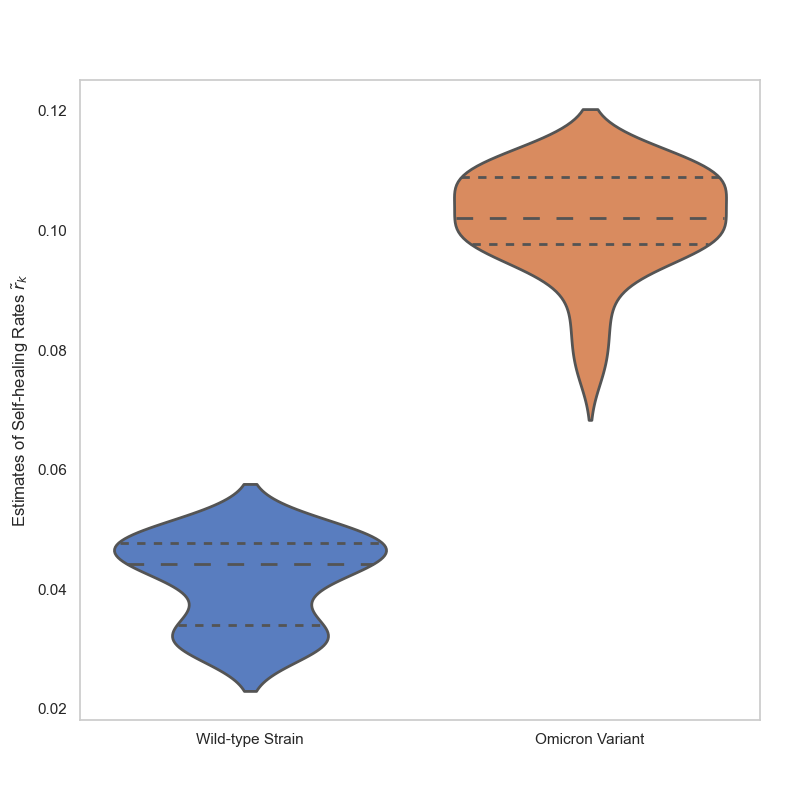}
                        \label{fig_rTld}
                \end{minipage}%
        }%
        \\
        \subfloat[{\scriptsize Time varying death rates of unaware infections $\tilde{d}_k$.}]{
                \begin{minipage}[h]{0.5\linewidth}
                        \centering
                        \includegraphics[scale=0.43]{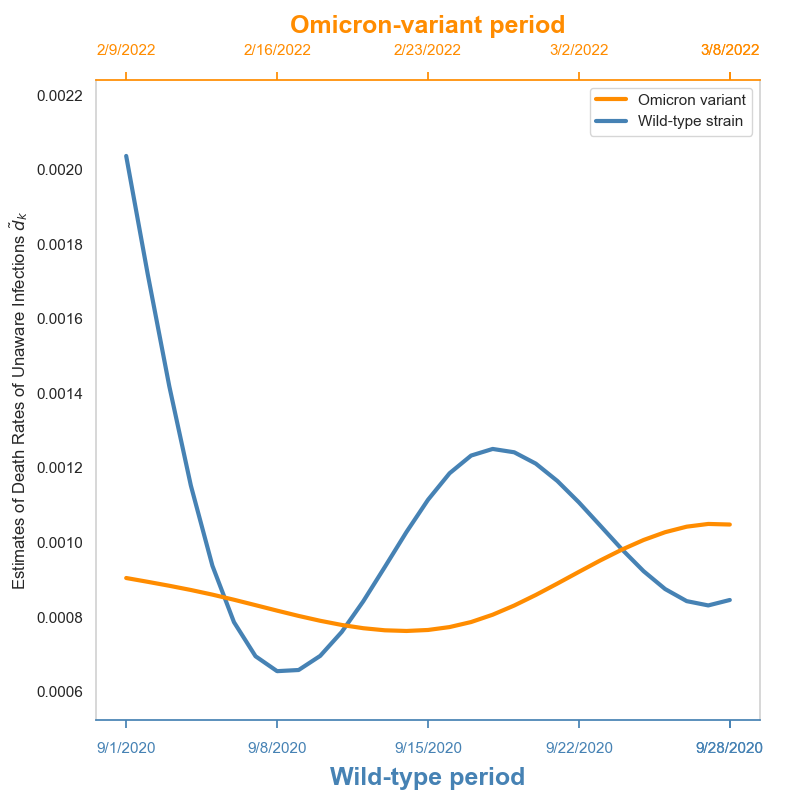}
                        \label{fig_dTld_line}
                \end{minipage}%
        }%
        \subfloat[{\scriptsize   Violin plot of the death rates of unaware infections.}]{
                \begin{minipage}[h]{0.5\linewidth}
                        \centering
                        \includegraphics[scale=0.43]{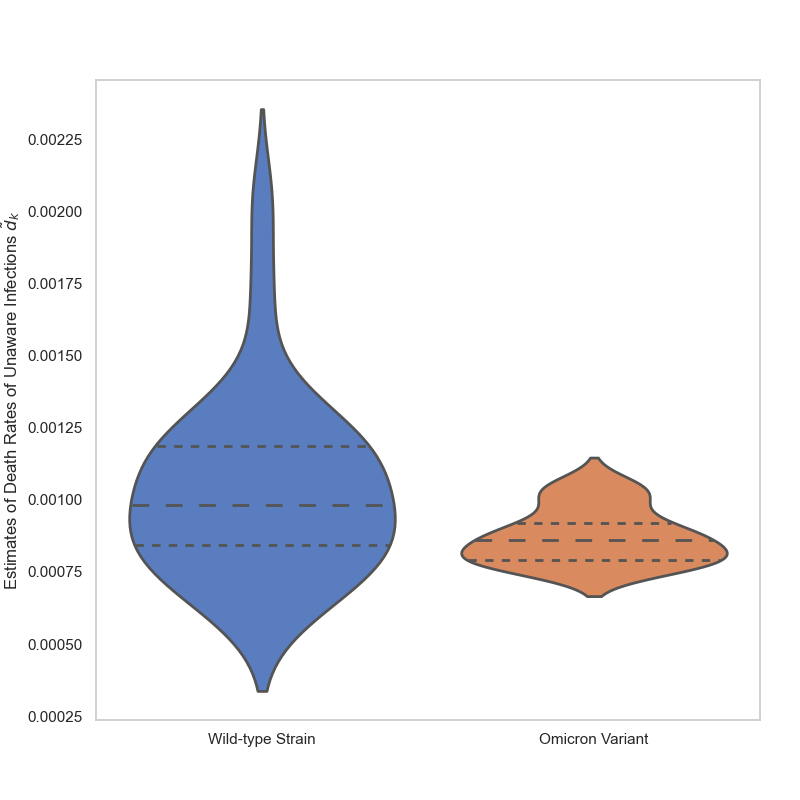}
                        \label{fig_dTld}
                \end{minipage}%
        }%
        \caption{Comparison of self-healing rates $\tilde{r}_k$ and death rates of unaware infections $\tilde{d}_k$ between wild-type and Omicron-variant periods.}
        \label{fig_timevaryingrates_comline}
\end{figure}

Figure \ref{fig_infe}  presents the estimates in the infection process. The left semicircle corresponds to the wild-type period, while the right semicircle corresponds to the Omicron-variant period. There is a noticeable discrepancy in the estimated values of $\alpha$. This divergence likely stems from distinct infection characteristics of the two strains. Therefore, developing strain-specific intervention strategies, tailored to these differences, would be beneficial.

Figure \ref{fig_conf} illustrates the comparison of parameters $\rho$, $\mu$, and $\psi$ between the wild-type and Omicron variants. A substantial difference is observed in the estimates of $\psi$, with a value of 0.7 during the wild-type period, which is more than double the 0.3 noted during the Omicron-variant period. In contrast, the estimates of $\hat\mu$ are stable across both periods, indicating that the effect of confirmed cases on subsequent infections remains consistent regardless of the strain.

Figure \ref{fig_timevaryingrates_comline}  demonstrate the evolution of $\tilde{r}_k$ and $\tilde{d}_k$ over time. These rates do {not} follow monotone trends   but display  variable patterns of increases and decreases. These variations   could be associated with strain characteristics, government policies and public behavior \cite{tian2021effects, badr2020association}.
 Specifically, in the Omicron-variant period, $\tilde{r}_k$ demonstrates a higher mean value, while $\tilde{d}_k$ shows a slightly lower mean value compared to the wild-type period. The findings suggest that the Omicron variant is less severe, aligning with the conclusions obtained in, for example  \cite{world2022severity, nealon2022omicron}. 

Finally, the \textit{self-healing rates} ($\tilde{r}_k$) and \textit{death rates of the unaware infections} ($\tilde{d}_k$) during different periods are shown in   Figures \ref{fig_rTld} and \ref{fig_dTld}, where the estimates of $\tilde{r}_k$ and $\tilde{d}_k$ exhibit distinct distributional characteristics  between different periods.

\subsection{{Unaware} Infections and {Uncertified} Deaths}
 Many COVID-19 analyses   focus on officially recorded cases and fatalities  due to data availability constraints, but estimating the numbers of undiagnosed infections ($U_k$) and unreported deaths ($D_{U,k}$) provides crucial insights.  Investigating these underreported segments can substantially refine our understanding of the epidemic's actual magnitude and transmission patterns.
\begin{figure}[t]
        \centering
        \subfloat[Wild-type period.]{
                \begin{minipage}[ht]{0.5\linewidth}
                        \centering
                        \includegraphics[scale=0.5]{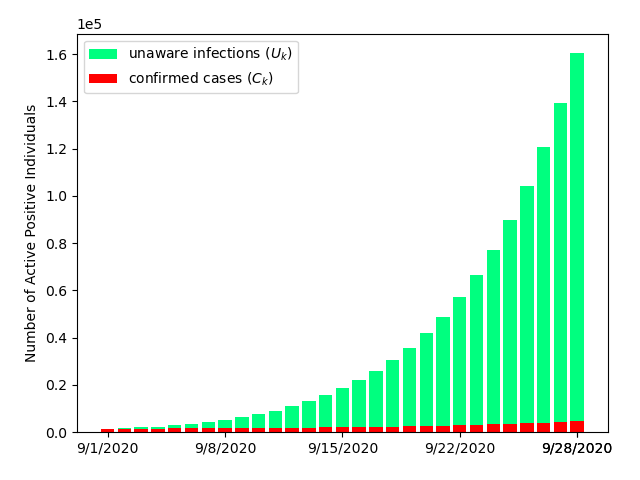}
                \end{minipage}\label{fig_z3_OR}
        }%
        \subfloat[Omicron-variant period.]{
                \begin{minipage}[ht]{0.5\linewidth}
                        \centering
                        \includegraphics[scale=0.5]{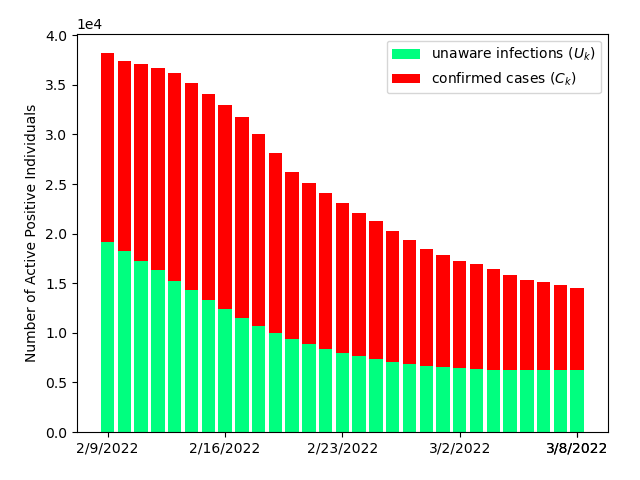}
                \end{minipage}\label{fig_z3_OM}
        }%
        \centering
        \caption{Comparison of the number of daily unaware infections and confirmed cases during the wild-type and Omicron-variant periods. In the wild-type period, the count of unaware infections significantly exceeds the number of confirmed cases (red). In the Omicron-variant period, the daily count of unaware infections averages less than 50\% of the confirmed cases.}
        \label{fig_z3_compare}
\end{figure}

\begin{figure}[t]
        \centering
        \subfloat[Wild-type period.]{
                \begin{minipage}[ht]{0.5\linewidth}
                        \centering
                        \includegraphics[scale=0.5]{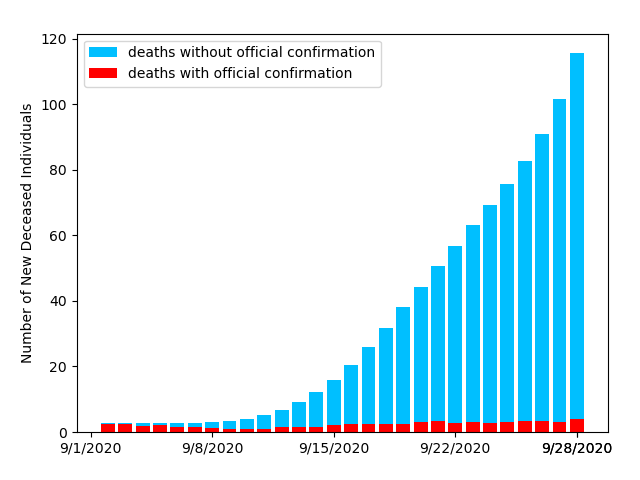}
                \end{minipage}\label{fig_z4_OR}
        }
        \subfloat[Omicron-variant period.]{
                \begin{minipage}[ht]{0.5\linewidth}
                        \centering
                        \includegraphics[scale=0.5]{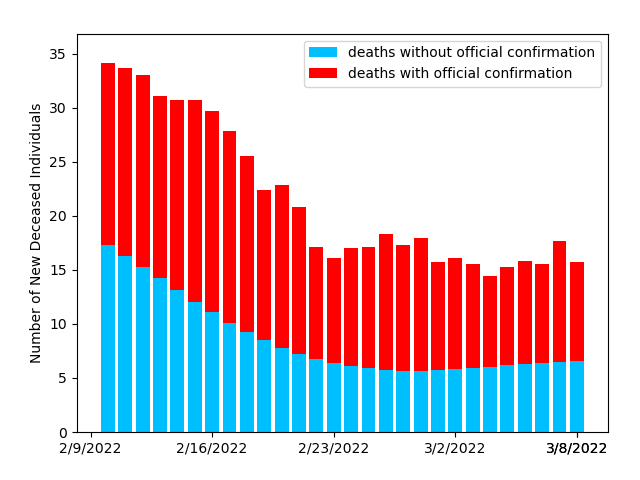}
                \end{minipage}\label{fig_z4_OM}
        }
        \centering
        \caption{Comparison of the number of daily new deaths without and with official confirmation during the wild-type and Omicron-variant periods. In the wild-type period, the daily count of new unconfirmed deaths markedly surpasses the confirmed deaths. In the Omicron-variant period, the daily count of new unconfirmed deaths still constitutes approximately 40\% of the confirmed deaths.}
        \label{fig_z4}
\end{figure}

Figure \ref{fig_z3_compare} gives  a comparison between undiagnosed infections ($U_k$) and confirmed cases ($C_k$) across the wild-type and Omicron-variant periods. During the wild-type period, the number of undiagnosed infections, represented in green, exhibited a marked increase, rising from 1,298 to 160,628. These figures highlight the significant presence of individuals unaware of their infection status. Notably, estimates of $U_k$ consistently surpassed those of $C_k$, underscoring a substantial underestimation of active infections. This discrepancy indicates that the true severity of the outbreak was likely obscured, contributing to prolonged epidemic conditions.


In contrast, during the Omicron-variant period, the estimated numbers of undiagnosed infections are significantly lower,   less than 50\% of the daily confirmed cases, and demonstrate a declining trend (from 19,127 to 6,282). This reduction suggests that government efforts to improve diagnostic capabilities and expand testing availability have been effective \cite{pitzer2021impact}.


Similarly, Figure \ref{fig_z4} presents the daily count of new deceased individuals, distinguishing between officially diagnosed deaths and those without official diagnoses across different periods. Current literature largely overlooks uncertified deaths \cite{yi2022characterizing, keller2022tracking}, while CURNDS is able to  estimate these unrecorded fatalities due to COVID-19 and evaluate their implications. According to Figure \ref{fig_z4_OR}, during the wild-type period, the average daily increase in uncertified deaths, shown in blue, is approximately 34, exceeding the number of confirmed deaths, shown in red, by a factor of 17.

During the Omicron-variant period (as depicted in Figure \ref{fig_z4_OM}), although uncertified deaths are significantly fewer than newly confirmed deaths, they still account for 40\% of the latter. These findings suggest that the actual death toll may be significantly higher than reported, highlighting the need to reassess the epidemic's impact and reallocate medical resources accordingly.

\subsection{Effective Reproduction Number}
\begin{figure}[h!]
        \centering
        \includegraphics[scale=0.5]{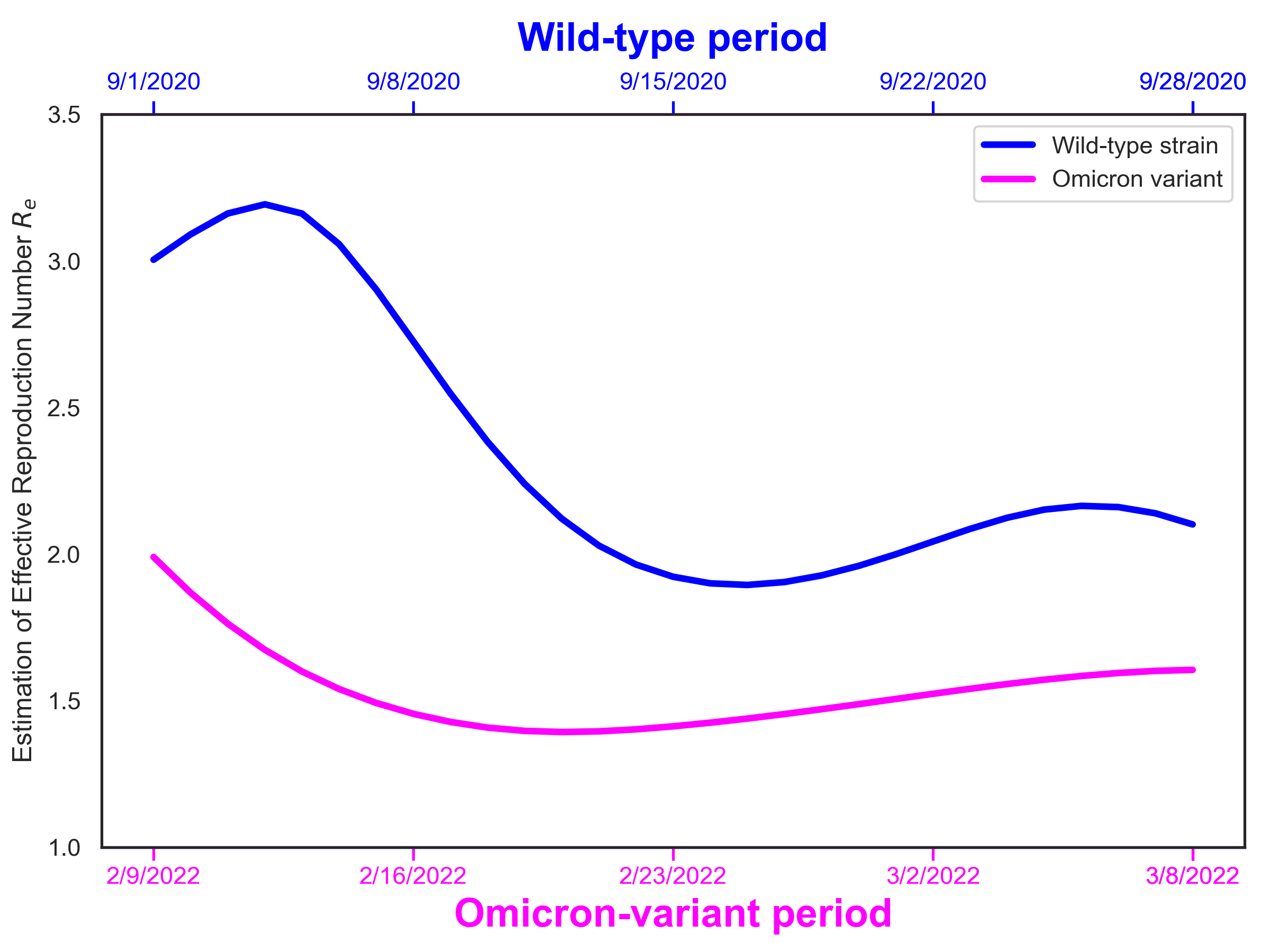}
        \centering
        \caption{Effective reproduction number curves,  $R_e(k)$,  for the wild-type and Omicron-variant periods exhibit distinct evolutionary patterns.  Notably, all values consistently exceed 1 throughout both periods.}
        \label{fig_Line_R_e}
\end{figure}

The \textit{effective reproduction number}, denoted as $R_e(k)$, measures  the extent of transmission in the presence of population immunity or interventions. It quantifies the average new infections each infected person generates during their contagious period. While   it is commonly acknowledged that self-healing rates ($\tilde{r}_k$), as well as death rates of unaware infections ($\tilde{d}_k$), should be incorporated to the calculation of $R_e(k)$, existing models   encounter difficulties in modeling and estimating numerous parameters.

Following the definitions   in \cite{hasan2022new} and utilizing parameter the estimates from our CURNDS model, we provide a   calculation for the effective reproduction number as follows \begin{equation}\label{ERN} \widehat{R}_e(k)=\frac{\widehat{N}_k}{P_0} \cdot \frac{\widehat{\varphi}}{\widehat{r}_k+\widehat{d}_k+\widehat{\tilde{r}}_k+\widehat{\tilde{d}}_k},
\end{equation}
where the involved parameters  are detailed in the Model Formulation section, 
 and $P_0$ represents the initial population size, which is generally a known quantity.

 Figure \ref{fig_Line_R_e} displays the effective reproduction numbers $\widehat{R}_e(k)$ for the wild-type and Omicron-variant periods, shown in blue and purple, respectively.   The purple curve consistently remains below the blue one. Importantly, the Quebec government, which did not implement COVID-19 vaccinations during the wild-type period, achieved over 84.6\% coverage during the Omicron period. This widespread vaccination likely contributed significantly to reducing virus transmission.  Nonetheless,   all   $\widehat{R}_e(k)$ estimates continue to exceed  1, indicating a sustained risk of epidemic persistence throughout both periods.



Moreover, the $R_e(k)$ curves highlight different ranges for the two periods. During the Omicron-variant period, the purple curve is stable, fluctuating between 1.4 and 2. In contrast, the blue curve representing the wild-type period shows significant variability, ranging from 1.9 to 3.2, and includes a pronounced decrease.  This decline may be attributed to the enforced implementation of interventions, such as social distancing, gathering restrictions, and traveling limitation \cite{howard2021evidence, tian2021effects}.
Overall, 
our method serves as a valuable tool for understanding evolving situations in the epidemic.


        \section{Summary}\label{Conclusion}

Traditional modeling and forecasting techniques face significant challenges from quarantinable, highly contagious diseases. The tracking and management of outbreaks are exacerbated by factors such as underreported mortality, asymptomatic transmission, delayed diagnoses, and changes in contact due to isolation. These issues underscore the urgent need for a model that prioritizes transmission based on contact levels, rather than solely focusing on symptom severity or hospitalization status.

The CURNDS model introduces several critical new population categories, such as ``unaware infections" ($\boldsymbol U$), individuals likely to spread the disease more widely than confirmed cases ($\boldsymbol C$) due to their frequent interactions and minimal isolation,  ``self-healed individuals" ($\boldsymbol S$), who recover without medical intervention---unlike officially recovered cases ($\boldsymbol R$), and it differentiates between documented deaths ($\boldsymbol D_C$) and unreported deaths ($\boldsymbol D_U$).

This framework offers a    comprehensive  mapping of transmission from initial infection to resolution, illustrated by transitions like  $\boldsymbol N\rightarrow \boldsymbol U\rightarrow \{\boldsymbol S, \boldsymbol C, \boldsymbol D_U\}$. It also incorporates non-linear infection rates and time-varying transmission dynamics to   account for non-uniform population mixing and disease progression affected  by prolonged isolation. The statistical modeling employs smooth nonparametric splines to characterize  rate curves and prevent overfitting. Overall, this new formulation boosts parameter robustness, predictive accuracy, and algorithm efficiency.

Our COVID-19 case studies validate CURNDS' effectiveness in real-world applications. The examination of power parameters  challenges the uniform mixing assumption prevalent in traditional epidemiological models. Our analysis of transmission dynamics offers critical insights into the spread of different COVID-19 strains, particularly in terms of infection and confirmation rates. A key finding  is the substantial presence of undiagnosed infections and unreported deaths during both the wild-type and Omicron phases of the pandemic, underscoring  critical gaps in current data collection and reporting methods.  In summary, CURNDS  offers a comprehensive framework for analyzing highly infectious, quarantinable diseases, improving  understanding of transmission dynamics and guiding effective interventions.      %

        \bibliographystyle{IEEEtran}
        \bibliography{IEEEabrv,reference}

\begin{thebibliography}{10}
\providecommand{\url}[1]{#1}
\csname url@samestyle\endcsname
\providecommand{\newblock}{\relax}
\providecommand{\bibinfo}[2]{#2}
\providecommand{\BIBentrySTDinterwordspacing}{\spaceskip=0pt\relax}
\providecommand{\BIBentryALTinterwordstretchfactor}{4}
\providecommand{\BIBentryALTinterwordspacing}{\spaceskip=\fontdimen2\font plus
\BIBentryALTinterwordstretchfactor\fontdimen3\font minus
  \fontdimen4\font\relax}
\providecommand{\BIBforeignlanguage}[2]{{%
\expandafter\ifx\csname l@#1\endcsname\relax
\typeout{** WARNING: IEEEtran.bst: No hyphenation pattern has been}%
\typeout{** loaded for the language `#1'. Using the pattern for}%
\typeout{** the default language instead.}%
\else
\language=\csname l@#1\endcsname
\fi
#2}}
\providecommand{\BIBdecl}{\relax}
\BIBdecl

\bibitem{Kermack1927}
W.~O. Kermack and A.~G. McKendrick, ``A contribution to the mathematical theory
  of epidemics,'' \emph{Proceedings of the Royal Society of London. Series A,
  Containing Papers of a Mathematical and Physical Character}, vol. 115, no.
  772, pp. 700--721, 1927.

\bibitem{hethcote1976qualitative}
H.~W. Hethcote, ``Qualitative analyses of communicable disease models,''
  \emph{Mathematical Biosciences}, vol.~28, no. 3-4, pp. 335--356, 1976.

\bibitem{he2020temporal}
X.~He, E.~H. Lau, P.~Wu, X.~Deng, J.~Wang, X.~Hao, Y.~C. Lau, J.~Y. Wong,
  Y.~Guan, X.~Tan \emph{et~al.}, ``Temporal dynamics in viral shedding and
  transmissibility of covid-19,'' \emph{Nature medicine}, vol.~26, no.~5, pp.
  672--675, 2020.

\bibitem{ferretti2020quantifying}
L.~Ferretti, C.~Wymant, M.~Kendall, L.~Zhao, A.~Nurtay, L.~Abeler-D{\"o}rner,
  M.~Parker, D.~Bonsall, and C.~Fraser, ``Quantifying sars-cov-2 transmission
  suggests epidemic control with digital contact tracing,'' \emph{science},
  vol. 368, no. 6491, p. eabb6936, 2020.

\bibitem{di2020dynamical}
P.~Di~Giamberardino, D.~Iacoviello, F.~Papa, and C.~Sinisgalli, ``Dynamical
  evolution of {COVID-19} in italy with an evaluation of the size of the
  asymptomatic infective population,'' \emph{IEEE Journal of Biomedical and
  Health Informatics}, vol.~25, no.~4, pp. 1326--1332, 2020.

\bibitem{biswas2020modelling}
M.~Biswas, M.~Islam, S.~Akter, S.~Mandal, M.~Khatun, S.~Samad, A.~Paul, and
  M.~Khatun, ``Modelling the effect of self-immunity and the impacts of
  asymptomatic and symptomatic individuals on {COVID-19} outbreak,''
  \emph{CMES-Computer Modeling in Engineering \& Sciences}, vol. 125, no.~3,
  pp. 1033--1060, 2020.

\bibitem{yi2022characterizing}
G.~Y. Yi, P.~Hu, and W.~He, ``Characterizing the {COVID-19} dynamics with a new
  epidemic model:
  susceptible-exposed-asymptomatic-symptomatic-active-removed,'' \emph{Canadian
  Journal of Statistics}, vol.~50, no.~2, pp. 395--416, 2022.

\bibitem{zhang2020prediction}
Y.~Zhang, C.~You, Z.~Cai, J.~Sun, W.~Hu, and X.-H. Zhou, ``Prediction of the
  covid-19 outbreak in {C}hina based on a new stochastic dynamic model,''
  \emph{Scientific reports}, vol.~10, no. 21522, 2020.

\bibitem{sayampanathan2021infectivity}
A.~A. Sayampanathan, C.~S. Heng, P.~H. Pin, J.~Pang, T.~Y. Leong, and V.~J.
  Lee, ``Infectivity of asymptomatic versus symptomatic {COVID-19},'' \emph{The
  Lancet}, vol. 397, no. 10269, pp. 93--94, 2021.

\bibitem{keller2022tracking}
J.~P. Keller, T.~Zhou, A.~Kaplan, G.~B. Anderson, and W.~Zhou, ``Tracking the
  transmission dynamics of {COVID-19} with a time-varying coefficient
  state-space model,'' \emph{Statistics in Medicine}, vol.~41, no.~15, pp.
  2745--2767, 2022.

\bibitem{yang2020modified}
Z.~Yang, Z.~Zeng, K.~Wang, S.-S. Wong, W.~Liang, M.~Zanin, P.~Liu, X.~Cao,
  Z.~Gao, Z.~Mai \emph{et~al.}, ``Modified {SEIR} and {AI} prediction of the
  epidemics trend of {COVID-19} in {C}hina under public health interventions,''
  \emph{Journal of Thoracic Disease}, vol.~12, no.~3, pp. 165--174, 2020.

\bibitem{mwalili2020seir}
S.~Mwalili, M.~Kimathi, V.~Ojiambo, D.~Gathungu, and R.~Mbogo, ``{SEIR} model
  for {COVID-19} dynamics incorporating the environment and social
  distancing,'' \emph{BMC Research Notes}, vol.~13, no. 352, 2020.

\bibitem{hassan2023mathematical}
M.~N. Hassan, M.~S. Mahmud, K.~F. Nipa, and M.~Kamrujjaman, ``Mathematical
  modeling and {COVID-19} forecast in {T}exas, {USA}: a prediction model
  analysis and the probability of disease outbreak,'' \emph{Disaster medicine
  and public health preparedness}, vol.~17, p. e19, 2023.

\bibitem{alzahrani2020forecasting}
S.~I. Alzahrani, I.~A. Aljamaan, and E.~A. Al-Fakih, ``Forecasting the spread
  of the covid-19 pandemic in {S}audi {A}rabia using {ARIMA} prediction model
  under current public health interventions,'' \emph{Journal of Infection and
  Public Health}, vol.~13, no.~7, pp. 914--919, 2020.

\bibitem{abuhasel2022analyzing}
K.~A. Abuhasel, M.~Khadr, and M.~M. Alquraish, ``Analyzing and forecasting
  {COVID-19} pandemic in the {K}ingdom of {S}audi {A}rabia using {ARIMA} and
  {SIR} models,'' \emph{Computational intelligence}, vol.~38, no.~3, pp.
  770--783, 2022.

\bibitem{Rustam2020}
F.~Rustam, A.~A. Reshi, A.~Mehmood, S.~Ullah, B.-W. On, W.~Aslam, and G.~S.
  Choi, ``Covid-19 future forecasting using supervised machine learning
  models,'' \emph{IEEE access}, vol.~8, pp. 101\,489--101\,499, 2020.

\bibitem{bezerra2020prediction}
A.~K.~L. Bezerra and {\'E}.~M.~C. Santos, ``Prediction the daily number of
  confirmed cases of {COVID-19} in {S}udan with {ARIMA} and {Holt Winter}
  exponential smoothing,'' \emph{International Journal of Development
  Research}, vol.~10, no.~08, pp. 39\,408--39\,413, 2020.

\bibitem{ribeiro2020short}
M.~H. D.~M. Ribeiro, R.~G. da~Silva, V.~C. Mariani, and L.~dos Santos~Coelho,
  ``Short-term forecasting {COVID-19} cumulative confirmed cases: Perspectives
  for {B}razil,'' \emph{Chaos, Solitons \& Fractals}, vol. 135, no. 109853,
  2020.

\bibitem{mishra2024multivariate}
S.~Mishra, T.~Singh, M.~Kumar, and Satakshi, ``Multivariate time series short
  term forecasting using cumulative data of coronavirus,'' \emph{Evolving
  Systems}, vol.~15, no.~3, pp. 811--828, 2024.

\bibitem{somyanonthanakul2022forecasting}
R.~Somyanonthanakul, K.~Warin, W.~Amasiri, K.~Mairiang, C.~Mingmalairak,
  W.~Panichkitkosolkul, K.~Silanun, T.~Theeramunkong, S.~Nitikraipot, and
  S.~Suebnukarn, ``Forecasting {COVID-19} cases using time series modeling and
  association rule mining,'' \emph{BMC medical research methodology}, vol.~22,
  no. 281, 2022.

\bibitem{Silva2020}
R.~G. da~Silva, M.~H. D.~M. Ribeiro, V.~C. Mariani, and L.~dos Santos~Coelho,
  ``Forecasting {Brazilian and American COVID-19} cases based on artificial
  intelligence coupled with climatic exogenous variables,'' \emph{Chaos,
  Solitons \& Fractals}, vol. 139, p. 110027, 2020.

\bibitem{xu2022forecasting}
L.~Xu, R.~Magar, and A.~B. Farimani, ``Forecasting {COVID-19} new cases using
  deep learning methods,'' \emph{Computers in biology and medicine}, vol. 144,
  p. 105342, 2022.

\bibitem{alali2022proficient}
Y.~Alali, F.~Harrou, and Y.~Sun, ``A proficient approach to forecast {COVID-19}
  spread via optimized dynamic machine learning models,'' \emph{Scientific
  Reports}, vol.~12, no.~1, p. 2467, 2022.

\bibitem{justus2018predicting}
D.~Justus, J.~Brennan, S.~Bonner, and A.~S. McGough, ``Predicting the
  computational cost of deep learning models,'' in \emph{2018 IEEE
  international conference on big data (Big Data)}.\hskip 1em plus 0.5em minus
  0.4em\relax IEEE, 2018, pp. 3873--3882.

\bibitem{wu2021interpretable}
H.~Wu, W.~Ruan, J.~Wang, D.~Zheng, B.~Liu, Y.~Geng, X.~Chai, J.~Chen, K.~Li,
  S.~Li \emph{et~al.}, ``Interpretable machine learning for {COVID-19}: An
  empirical study on severity prediction task,'' \emph{IEEE Transactions on
  Artificial Intelligence}, vol.~4, no.~4, pp. 764--777, 2021.

\bibitem{zheng2021health}
W.~Zheng, F.~K{\"a}mpfen, and Z.~Huang, ``Health-seeking and diagnosis delay
  and its associated factors: a case study on {COVID-19} infections in
  {S}haanxi province, {C}hina,'' \emph{Scientific Reports}, vol.~11, no.~1, p.
  17331, 2021.

\bibitem{anderson1991}
R.~M. Anderson and R.~M. May, \emph{{Infectious Diseases of Humans: Dynamics
  and Control}}.\hskip 1em plus 0.5em minus 0.4em\relax Oxford University
  Press, 1991.

\bibitem{kwuimy2020nonlinear}
C.~Kwuimy, F.~Nazari, X.~Jiao, P.~Rohani, and C.~Nataraj, ``Nonlinear dynamic
  analysis of an epidemiological model for covid-19 including public behavior
  and government action,'' \emph{Nonlinear Dynamics}, vol. 101, pp. 1545--1559,
  2020.

\bibitem{Berx2022Epid}
J.~Berx and J.~O. Indekeu, ``Epidemic processes with vaccination and immunity
  loss studied with the {BLUES} function method,'' \emph{Physica A: Statistical
  Mechanics and its Applications}, vol. 590, p. 126724, 2022.

\bibitem{chen2020covid}
S.~Chen, J.~Yang, W.~Yang, C.~Wang, and T.~B{\"a}rnighausen, ``Covid-19 control
  in {C}hina during mass population movements at new year,'' \emph{The Lancet},
  vol. 395, no. 10226, pp. 764--766, 2020.

\bibitem{gong2023sars}
W.~Gong, S.~Parkkila, X.~Wu, and A.~Aspatwar, ``Sars-cov-2 variants and
  {COVID-19} vaccines: Current challenges and future strategies,''
  \emph{International reviews of immunology}, vol.~42, no.~6, pp. 393--414,
  2023.

\bibitem{hawkins2004problem}
D.~M. Hawkins, ``The problem of overfitting,'' \emph{Journal of chemical
  information and computer sciences}, vol.~44, no.~1, pp. 1--12, 2004.

\bibitem{lai2020effect}
S.~Lai, N.~W. Ruktanonchai, L.~Zhou, O.~Prosper, W.~Luo, J.~R. Floyd,
  A.~Wesolowski, M.~Santillana, C.~Zhang, X.~Du \emph{et~al.}, ``Effect of
  non-pharmaceutical interventions to contain {COVID-19} in {China},''
  \emph{nature}, vol. 585, no. 7825, pp. 410--413, 2020.

\bibitem{wahba1990spline}
G.~Wahba, \emph{Spline models for observational data}.\hskip 1em plus 0.5em
  minus 0.4em\relax SIAM, 1990.

\bibitem{hastie2009elements}
T.~Hastie, R.~Tibshirani, J.~H. Friedman, and J.~H. Friedman, \emph{The
  elements of statistical learning: data mining, inference, and
  prediction}.\hskip 1em plus 0.5em minus 0.4em\relax Springer, 2009, vol.~2.

\bibitem{Hasell2020}
J.~Hasell, E.~Mathieu, D.~Beltekian, B.~Macdonald, C.~Giattino,
  E.~Ortiz-Ospina, M.~Roser, and H.~Ritchie, ``A cross-country database of
  {COVID-19} testing,'' \emph{Scientific data}, vol.~7, no.~1, p. 345, 2020.

\bibitem{Zhan2020Modeling}
C.~Zhan, C.~K. Tse, Y.~Fu, Z.~Lai, and H.~Zhang, ``Modeling and prediction of
  the 2019 coronavirus disease spreading in {C}hina incorporating human
  migration data,'' \emph{Plos one}, vol.~15, no.~10, p. e0241171, 2020.

\bibitem{Wachter2006}
A.~W{\"a}chter and L.~T. Biegler, ``On the implementation of an interior-point
  filter line-search algorithm for large-scale nonlinear programming,''
  \emph{Mathematical Programming}, vol. 106, no.~1, pp. 25--57, 2006.

\bibitem{yonar2020modeling}
H.~Yonar, A.~Yonar, M.~A. Tekindal, and M.~Tekindal, ``Modeling and forecasting
  for the number of cases of the {COVID}-19 pandemic with the curve estimation
  models, the {Box-Jenkins} and exponential smoothing methods,'' \emph{Eurasian
  Journal of Medicine and Oncology}, vol.~4, no.~2, pp. 160--165, 2020.

\bibitem{Guleryuz2021forecasting}
D.~Guleryuz, ``Forecasting outbreak of {COVID-19} in {T}urkey; comparison of
  {Box-Jenkins}, {Brown}’s exponential smoothing and long short-term memory
  models,'' \emph{Process Safety and Environmental Protection}, vol. 149, pp.
  927--935, 2021.

\bibitem{canadagov}
{Government of Canada}, ``{COVID-19} epidemiology update: Key updates,''
  \url{https://health-infobase.canada.ca/covid-19/}, 2022.

\bibitem{saraiva2023hierarchical}
E.~F. Saraiva, L.~Sauer, and C.~A. d.~B. Pereira, ``A hierarchical {B}ayesian
  approach for modeling the evolution of the 7-day moving average of the number
  of deaths by {COVID}-19,'' \emph{Journal of Applied Statistics}, vol.~50,
  no.~10, pp. 2194--2208, 2023.

\bibitem{holt2004forecasting}
C.~C. Holt, ``Forecasting seasonals and trends by exponentially weighted moving
  averages,'' \emph{International journal of forecasting}, vol.~20, no.~1, pp.
  5--10, 2004.

\bibitem{howard2021evidence}
J.~Howard, A.~Huang, Z.~Li, Z.~Tufekci, V.~Zdimal, H.-M. Van Der~Westhuizen,
  A.~Von~Delft, A.~Price, L.~Fridman, L.-H. Tang \emph{et~al.}, ``An evidence
  review of face masks against {COVID}-19,'' \emph{Proceedings of the National
  Academy of Sciences}, vol. 118, no.~4, p. e2014564118, 2021.

\bibitem{tian2021effects}
T.~Tian, J.~Tan, W.~Luo, Y.~Jiang, M.~Chen, S.~Yang, C.~Wen, W.~Pan, and
  X.~Wang, ``The effects of stringent and mild interventions for coronavirus
  pandemic,'' \emph{Journal of the American Statistical Association}, vol. 116,
  no. 534, pp. 481--491, 2021.

\bibitem{badr2020association}
H.~S. Badr, H.~Du, M.~Marshall, E.~Dong, M.~M. Squire, and L.~M. Gardner,
  ``Association between mobility patterns and {COVID-19} transmission in the
  {USA}: a mathematical modelling study,'' \emph{The Lancet Infectious
  Diseases}, vol.~20, no.~11, pp. 1247--1254, 2020.

\bibitem{world2022severity}
W.~H. Organization, \emph{Severity of disease associated with Omicron variant
  as compared with Delta variant in hospitalized patients with suspected or
  confirmed SARS-CoV-2 infection}.\hskip 1em plus 0.5em minus 0.4em\relax World
  Health Organization, 2022.

\bibitem{nealon2022omicron}
J.~Nealon and B.~J. Cowling, ``Omicron severity: milder but not mild,''
  \emph{The Lancet}, vol. 399, no. 10323, p. 412, 2022.

\bibitem{pitzer2021impact}
V.~E. Pitzer, M.~Chitwood, J.~Havumaki, N.~A. Menzies, S.~Perniciaro, J.~L.
  Warren, D.~M. Weinberger, and T.~Cohen, ``The impact of changes in diagnostic
  testing practices on estimates of {COVID}-19 transmission in the {U}nited
  {S}tates,'' \emph{American Journal of Epidemiology}, vol. 190, no.~9, pp.
  1908--1917, 2021.

\bibitem{hasan2022new}
A.~Hasan, H.~Susanto, V.~Tjahjono, R.~Kusdiantara, E.~Putri, N.~Nuraini, and
  P.~Hadisoemarto, ``A new estimation method for {COVID-19} time-varying
  reproduction number using active cases,'' \emph{Scientific Reports}, vol.~12,
  no.~1, p. 6675, 2022.

\end{thebibliography}
\end{document}